\documentclass[12pt,letterpaper]{article}
\title{Symmetries of supergravity black holes}
\author{
David D. K. Chow
}
\date{}

\usepackage{amsmath}
\usepackage{amsfonts}
\usepackage{amssymb}
\usepackage{amscd}
\usepackage{latexsym}

\makeatletter
\@addtoreset{equation}{section}
\makeatother

\newcommand{\ben}{\begin{equation}}
\newcommand{\een}{\end{equation}}
\newcommand{\bea}{\setlength\arraycolsep{2pt} \begin{eqnarray}}
\newcommand{\eea}{\end{eqnarray}}
\newcommand{\nnr}{\nonumber \\}
\newcommand{\pd}{\partial}
\newcommand{\ud}{\textrm{d}}
\newcommand{\ui}{\textrm{i}}
\newcommand{\ue}{\textrm{e}}
\newcommand{\uU}{\textrm{U}}
\newcommand{\uSO}{\textrm{SO}}
\newcommand{\uSU}{\textrm{SU}}
\newcommand{\te}{\tilde{e}}

\begin{document}
\begin{titlepage}
\begin{flushright}
DAMTP-2008-103\\
MIFP-08-27
\end{flushright}
\vspace*{100pt}
\begin{center}
{\bf \Large{Symmetries of supergravity black holes}}\\
\vspace{50pt}
\large{David D. K. Chow}
\end{center}
\begin{center}
{\it Department of Applied Mathematics and Theoretical Physics, University of Cambridge,\\
Centre for Mathematical Sciences, Wilberforce Road, Cambridge CB3 0WA, UK}\\
\vspace*{12pt}
and\\
\vspace*{12pt}
{\it George P. \& Cynthia W. Mitchell Institute for Fundamental Physics and Astronomy,\\
Texas A\&M University, College Station, TX 77843-4242, USA}\footnote{Present address}\\
{\tt chow@physics.tamu.edu}\\
\vspace{50pt}
{\bf Abstract\\}
\end{center}
We investigate Killing tensors for various black hole solutions of supergravity theories.  Rotating black holes of an ungauged theory, toroidally compactified heterotic supergravity, with NUT parameters and two $\uU (1)$ gauge fields are constructed.  If both charges are set equal, then the solutions simplify, and then there are concise expressions for rank-2 conformal Killing--St\"{a}ckel tensors.  These are induced by rank-2 Killing--St\"{a}ckel tensors of a conformally related metric that possesses a separability structure.  We directly verify the separation of the Hamilton--Jacobi equation on this conformally related metric, and of the null Hamilton--Jacobi and massless Klein--Gordon equations on the ``physical'' metric.  Similar results are found for more general solutions; we mainly focus on those with certain charge combinations equal in gauged supergravity, but also consider some other solutions.
\end{titlepage}

\newpage

\tableofcontents

\newpage


\section{Introduction}


A higher-dimensional analogue of the Kerr--Newman solution, i.e.~a charged and rotating exact black hole solution of the Einstein--Maxwell system in more than 4 spacetime dimensions, is not known.  We do know a higher-dimensional Kerr solution, namely the Myers--Perry solution \cite{bhhigherdim}, which describes an uncharged rotating black hole; and we also know a higher-dimensional Reissner--Nordstr\"{o}m solution \cite{bhhigherdim}, which describes a charged non-rotating black hole.  Recent work on a higher-dimensional Kerr--Newman solution includes the slow rotation limit \cite{slowrotcbh5d, rotbhhdEMgrav} and numerical work \cite{5dcrotbh, crotbhoddd}.  Despite these various approaches, a general exact solution in the context of Einstein--Maxwell theory has so far proved elusive.

However, the main motivation for studying higher-dimensional black holes is from string theory and M-theory, for which the relevant gravitational theories are generally not Einstein--Maxwell theories, but instead supergravity theories.  There has been more success in studying charged and rotating exact black hole solutions of these supergravity theories.  This is because ungauged supergravity theories possess global symmetries, giving a mechanical solution generating technique that produces a charged solution from an uncharged solution.  Starting from the Myers--Perry solution, which is an asymptotically flat uncharged rotating black hole in spacetime dimension $D \geq 4$, one can generate asymptotically flat charged rotating black holes that carry various numbers of abelian $\uU (1)$ charges.  Examples of solutions obtained in this way are: the 4-charge Cveti\v{c}--Youm solution in $D = 4$ \cite{entropycrotbhst, crotbh4d}, the 3-charge Cveti\v{c}--Youm solution in $D = 5$ \cite{rot5dbhhet}, and the 2-charge Cveti\v{c}--Youm solution in $D \geq 4$ \cite{nearBPSsat}.  The 2-charge Cveti\v{c}--Youm solution is a special case of the 4- and 3-charge solutions in $D = 4$ and $D = 5$ respectively, and so, as emphasised in \cite{equalcharge}, underlies solutions in a variety of dimensions.

However, in the AdS/CFT correspondence \cite{AdSCFT, gaugecorncst, AdShol, AdSCFTPhysRep}, it is instead asymptotically AdS solutions of gauged supergravity theories that are of interest.  These theories do not possess the global symmetries of their ungauged counterparts, so there is no similar charging procedure, making the construction of charged and rotating black hole solutions less straightforward.  Rather than systematic construction, some degree of guesswork is required.  Nevertheless, such guesswork can be minimized and well-motivated in certain special cases, and has succeeded.  All known constructions possess ``symmetry'', in a different sense of the word to the global symmetries that we have mentioned; these solutions involve Killing vectors, tensors or spinors: symmetries specific to certain solutions of a particular theory.

Work has concentrated on gauged supergravity theories in $D = 4, 5, 6, 7$.  All known solutions truncate to the Cartan subgroup of the full gauge group, namely $\uU (1)^4 \subset \uSO (8)$ in $D = 4$, $\uU (1)^3 \subset \uSO (6)$ in $D = 5$, $\uU (1) \subset \uSU (2)$ in $D = 6$, and $\uU (1)^2 \subset \uSO (5)$ in $D = 7$.  In $D = 2n + \varepsilon$ dimensions, $\varepsilon = 0, 1$, the rotation group $\uSO (D-1)$ has rank $\lfloor (D-1)/2 \rfloor = n - 1 + \varepsilon$, which is the number of independent angular momenta describing rotation in orthogonal 2-planes.  There should be general black hole solutions of these gauged supergravity theories with the maximum number of independent angular momenta and $\uU (1)$ charges, plus a mass parameter and an arbitrary gauge-coupling constant.  Only for the $D = 6$, $\uSU (2)$ gauged supergravity theory is such a solution known \cite{crotbh6}.  Such general solutions would include: in the ungauged limit, cases of the various Cveti\v{c}--Youm solutions \cite{entropycrotbhst, crotbh4d, rot5dbhhet, nearBPSsat}; in the uncharged limit, the higher-dimensional Kerr--AdS solution \cite{genKerrdS, rotbhhigherdim} (see also \cite{genKNUTAdSalld} for the presentation that we use); and in the non-rotating limit, static and spherically symmetric charged black holes in gauged supergravity, which are known in 4 \cite{AdSbhgN=8sugra}, 5 \cite{bh5dAdSN=2sugra}, 6 \cite{6dsugraIIA} and 7 \cite{phasesRcbh, bhmemAdS7} dimensions.

There are 3 simplification strategies to obtain exact black hole solutions that are charged and rotating in various gauged supergravity theories:
\begin{enumerate}
\item Restricting to supersymmetric solutions.
\item Setting all of the angular momenta equal.
\item Setting certain combinations of charges equal.
\end{enumerate}
As we shall explain, each of these is or seems to be associated with some type of extra symmetry, respectively:
\begin{enumerate}
\item The existence of a Killing spinor.
\item The existence of extra Killing vectors.
\item The existence of special rank-2 Killing--St\"{a}ckel tensors.
\end{enumerate}

Underlying the first approach of supersymmetric solutions are classifications based on writing solutions in a canonical form adapted to a Killing spinor.  The classification depends on the particular theory, and has been carried out for ungauged and gauged theories in various dimensions.  Since the classification becomes more implicit as the dimension is increased, this approach is most fruitful in low dimensions.  For example, building on the classification for minimal 5-dimensional gauged supergravity \cite{5dgaugedsugra}, supersymmetric AdS$_5$ black holes have been constructed \cite{susyAdS5bh, gensusyAdS5bh, susymcAdS5}.  Nevertheless, some supersymmetric AdS black holes have been found as limits of non-extremal solutions.

Underlying the second approach of equal angular momenta, for $D \geq 5$, is an enhancement of the rotational symmetry group.  If certain 2-planes have the same rotation parameter, then the symmetry between these planes is realised by extra Killing vectors that ``rotate'' these 2-planes amongst each other.  Such Killing vectors for the higher-dimensional Kerr--AdS metric are explicitly given in \cite{sepHJKGKdS}.  The rotational symmetry group in $D = 2n + \varepsilon$ dimensions is generally $\uU (1)^{n - 1 + \varepsilon}$, but if all of the angular momenta are equal and non-zero, then it is enhanced to $\uU ( n - 1 + \varepsilon )$.  Much of the angular dependence of the metric can then be packaged into a Fubini--Study metric on $\mathbb{CP}^{n-2+\varepsilon}$ \cite{genKerrdS, rotbhhigherdim}, which is $\uU (n-1+\varepsilon)$-invariant.  This extra symmetry continues to hold with the inclusion of charges, leading to the construction of gauged supergravity black holes in 5 \cite{crotbh5dU13} and 7 \cite{crotbh7d} dimensions.

The third approach of setting certain combinations of charges equal is what we mainly consider in this paper.  There are certain vielbeins through which the solutions can be written rather simply, generalizing the manner in which the higher-dimensional Kerr--AdS metric is presented in \cite{genKNUTAdSalld}.  We can present the solution so that certain coordinates and parameters appear on an equal footing, i.e.~there are discrete symmetries that permute coordinates and parameters (yet another sense of the word ``symmetry'').  The approach has been used to simplify the 2-charge Cveti\v{c}--Youm solution of ungauged supergravity when both charges are equal \cite{equalcharge}, from which one can conjecture fairly tight ansatzes to find gauged generalizations.  Using this simplification, charged and rotating black holes in gauged supergravity were constructed in 4 \cite{crotbh4d}, 5 \cite{5dgsugrabhind, bh5dgsugra, nonextremrotbh5dgsugra, newrotbh5d}, 6 \cite{crotbh6} and 7 \cite{equalcharge} dimensions.

The equal charge simplification has applications beyond providing conjectures for new solutions.  Its vielbeins are advantageous for calculations; for example, to compute the curvature to check explicitly that the Kerr--NUT--AdS solution in any dimension solves the Einstein equation \cite{KNdScurv}. Furthermore, many black hole solutions in higher dimensions possess various symmetric or antisymmetric Killing or conformal Killing tensors.  There can be simple expressions for symmetric (conformal) Killing--St\"{a}ckel tensors in terms of the vielbeins; conversely, one can regard the simplest choice of vielbeins as being adapted to such tensors.  For Killing tensors of the higher-dimensional Kerr--NUT--AdS metric, see \cite{hidsymhdKNUTAdS, integgeo, KYtKt, cogmhdbh} (see also \cite{hdbhhidsymsepvar, KubiznakPhD} for a review); in this uncharged case, the symmetric (conformal) Killing--St\"{a}ckel tensors can be obtained from antisymmetric (conformal) Killing--Yano tensors.  The existence of such tensors underlies the separability of, for example, the Hamilton--Jacobi equation (HJE) for geodesic motion and the Klein--Gordon equation (KGE).  For the higher-dimensional Kerr--NUT--AdS metric, the separation of various equations has been explicitly carried out \cite{sepHJKGKNUTAdS, sepDiracKNUTdS, stationaryKNUTAdS, parallelKNUTAdS, parallelnullgeo}.

This paper further explores a variety of black hole solutions in supergravity theories, studying geometrical aspects of known solutions.  In particular, we study their (conformal) Killing tensors, which leads to the separation of equations such as the HJE.

The outline of this paper is as follows.  In Section 2, we construct a higher-dimensional Kerr--NUT solution with two equal $\uU (1)$ charges in ungauged supergravity, which generalizes the 2-charge Cveti\v{c}--Youm solution, with both charges equal, to include NUT parameters.  It can be viewed as a solution of toroidally compactified heterotic supergravity and in certain dimensions as a solution of the ungauged limit of gauged supergravity theories.  This equal charge simplification is a generalization of that in \cite{equalcharge}, which dealt with the 2-charge Cveti\v{c}--Youm solution with both charges equal, which is in turn a generalization of the way the Myers--Perry metric was presented in \cite{genKNUTAdSalld}.  In Section 3, we consider the hidden symmetries and separability properties of this solution, showing that there is a conformally related metric that possesses Killing--St\"{a}ckel tensors, inducing conformal Killing--St\"{a}ckel tensors for the ``physical'' Einstein frame metric.  There is a separability structure, and we also directly verify separation of the HJE for the conformally related metric and the massless KGE for the ``physical'' metric.  In Section 4, we consider black holes in $D = 4, 5, 6, 7$ gauged supergravity theories that have certain combinations of charges set equal, and in Section 5, we consider certain black holes in ungauged supergravity theories, including the 2-charge and 4-charge Cveti\v{c}--Youm solutions in 4 dimensions.  For all these examples, we obtain various (conformal) Killing--St\"{a}ckel tensors, unifying some that have appeared in the literature; these tensors are induced by Killing--St\"{a}ckel tensors of a conformally related metric.  We conclude in Section 6.  The Appendix records a general 2-charge Kerr--NUT solution in higher dimensions.


\section{Higher-dimensional charged Kerr--NUT}


We first construct a higher-dimensional charged Kerr--NUT solution, which generalizes the 2-charge Cveti\v{c}--Youm solution \cite{nearBPSsat} to include NUT parameters.  Even though we are largely not concerned with the NUT parameters, it is easier for general calculations and for context to include them, because they appear on a symmetrical footing to the mass parameter.  The construction involves charging up the uncharged higher-dimensional Kerr--NUT solution using a solution generating technique.


\subsection{Charging procedure}


We consider a Lagrangian that appears as a truncation of the bosonic sector of various supergravity theories, for example of heterotic supergravity compactified on a torus.  It also appears as the ungauged limit of truncations of certain gauged supergravity theories.  The Lagrangian in $D \geq 4$ spacetime dimensions is
\ben
\mathcal{L}_D = R \star 1 - \frac{1}{2} \sum_{i=1}^2 \star \ud
\varphi_i \wedge \ud \varphi_i - \frac{1}{2} \sum_{I=1}^2 X_I^{-2}
\star F_{(2)}^I \wedge F_{(2)}^I - \frac{1}{2} X_1^{-2} X_2^{-2} \star
H_{(3)} \wedge H_{(3)} ,
\label{Lagrangian}
\een
where
\bea
&& X_1 = \ue ^{- \varphi_1/\sqrt{2(D-2)} - \varphi_2/\sqrt{2}} , \quad
X_2 = \ue ^{- \varphi_1/\sqrt{2(D-2)} + \varphi_2/\sqrt{2}} , \nnr
&& F_{(2)}^I = \ud A_{(1)}^I , \quad H_{(3)} = \ud B_{(2)} -
\tfrac{1}{2} A_{(1)}^1 \wedge \ud A_{(1)}^2 - \tfrac{1}{2} A_{(1)}^2
\wedge \ud A_{(1)}^1 .
\eea

It can be more convenient to dualize the 2-form potential and
3-form field strength in favour of a $(D - 4)$-form potential and
$(D - 3)$-form field strength.  The dual Lagrangian is
\bea
\mathcal{L}_D & = & R \star 1 - \frac{1}{2} \sum_{i=1}^2 \star \ud
\varphi_i \wedge \ud \varphi_i - \frac{1}{2} \sum_{I=1}^2 X_I^{-2} \star F_{(2)}^I
\wedge F_{(2)}^I - \frac{1}{2} X_1^2 X_2^2 \star F_{(D - 3)} \wedge F_{(D - 3)} \nnr
&& + (-1)^{D-1} F_{(2)}^1 \wedge F_{(2)}^2 \wedge A_{(D - 4)} ,
\eea
where $F_{(2)}^I = \ud A_{(1)}^I$ and
\ben
F_{(D - 3)} = \ud A_{(D - 4)} = X_1^{-2} X_2^{-2} \star H_{(3)} . 
\label{dualF}
\een

The Lagrangian (\ref{Lagrangian}) may be obtained from reducing on a circle the $(D + 1)$-dimensional ``bosonic string theory'' Lagrangian
\ben
\mathcal{L}_{D + 1} = R \star 1 - \tfrac{1}{2} \star \ud \phi_1 \wedge \ud \phi_1 - \tfrac{1}{2} \ue ^{2 \sqrt{2/(D-1)} \phi_1} \star H_{(3)} \wedge H_{(3)} .
\een
Reduction of the Einstein--Hilbert term gives a scalar $\phi_2$,
i.e.~a dilaton, and a 1-form potential $A_{(1)}^2$.  Reduction of
the 3-form field strength gives another 1-form potential
$A_{(1)}^1$.  The metric reduction ansatz is
\ben
\ud s_{D + 1}^2 = \ue ^{-\sqrt{2 (D - 2)/(D - 1)} \phi_2} (\ud z + A_{(1)}^2)^2 + \ue ^{\sqrt{2 /(D - 1)(D - 2)} \phi_2} \ud s_D^2 ,
\een
and the 3-form field strength decomposes as
\ben
\hat{H}_{(3)} = \ue ^{- 3 \phi_2 / \sqrt{2(D-1)(D-2)}} H_{(3)} + \ue ^{(D-4) \phi_2 / \sqrt{2(D-1)(D-2)}} F_{(2)}^1 \wedge (\ud z + A_{(1)}^2) ,
\een
here denoting the $(D + 1)$-dimensional 3-form field strength with a hat for clarity.  The Lagrangian in the form (\ref{Lagrangian}) is then recovered by rotating the 2 scalars, defining
\ben
\begin{pmatrix}
\varphi_1 \\
\varphi_2
\end{pmatrix}
= \frac{1}{\sqrt{D-1}}
\begin{pmatrix}
\sqrt{D-2} & -1 \\
1 & \sqrt{D-2}
\end{pmatrix}
\begin{pmatrix}
\phi_1 \\
\phi_2
\end{pmatrix} .
\een

We may therefore systematically charge up a solution of the vacuum Einstein equation in $D$ dimensions, which solves the field equations of the $D$-dimensional Lagrangian, lifting to $D+1$ dimensions, performing a Lorentz boost along the extra coordinate,
\ben
\begin{pmatrix}
t \\
z
\end{pmatrix}
\rightarrow
\begin{pmatrix}
\cosh \delta_1 & \sinh \delta_1 \\
\sinh \delta_1 & \cosh \delta_1
\end{pmatrix}
\begin{pmatrix}
t \\
z
\end{pmatrix} ,
\een
and then reducing back to $D$ dimensions.  The second charge is introduced by swapping $A_{(1)}^1$ and $A_{(1)}^2$ before repeating the procedure with a second boost parameter $\delta_2$.

Starting with the Myers--Perry metric \cite{bhhigherdim}, which describes an uncharged black hole with all angular momenta independent, one can obtain the 2-charge Cveti\v{c}--Youm solution \cite{nearBPSsat} (although a different method was used originally).  We shall generalize to include further NUT parameters, because then the symmetries of the solution become more apparent.


\subsection{Higher-dimensional Kerr--NUT}


The uncharged solution that we consider is the higher-dimensional Kerr--NUT metric, which is the zero cosmological constant limit of the Kerr--NUT--AdS metric \cite{genKNUTAdSalld}.  It is a generalization of the Myers--Perry metric \cite{bhhigherdim}: as well as a mass parameter, the solution contains a number of NUT parameters.  By appropriate analytic continuation, these mass and NUT parameters may be placed on a symmetrical footing.


\subsubsection{Even dimensions $D = 2n$}


The analytically continued Kerr--NUT metric in even dimensions is
\ben
\ud s^2 = \sum_{\mu = 1}^n \left( \frac{X_\mu}{U_\mu}
  \mathcal{A}_\mu^2 + \frac{U_\mu}{X_\mu} \, \ud x_\mu^2 \right) ,
\label{KNeven}
\een
where
\bea
&& U_\mu = \sideset{}{'} \prod_{\nu = 1}^n (x_\nu^2 - x_\mu^2) , \quad
X_\mu = - \prod_{k=1}^{n-1} (a_k^2 - x_\mu^2) + 2 m_\mu x_\mu , \nnr
&& \tilde{\gamma}_i = \prod_{\nu = 1}^n (a_i^2 -
  x_\nu^2) , \quad \mathcal{A}_\mu = \ud t -  \sum_{i=1}^{n-1} \frac{\tilde{\gamma}_i}{a_i^2 - x_\mu^2} \, \ud
  \tilde{\phi}_i , \quad \tilde{\phi}_i = \frac{\phi_i}{a_i \prod^\prime{_{k=1}^{n-1}} (a_i^2 - a_k^2)} .
\label{KNeven2}
\eea
The notation $\prod ^\prime$ indicates that we exclude from a product the factor that vanishes.  $\phi_i$ are canonically normalized Boyer--Lindquist coordinates for the azimuthal angles.  The Lorentzian metric is recovered by the analytic continuations $x_n = \ui r$ and $m_n = - \ui m$; $r$ is a radial coordinate and $m$ is a mass parameter.  These same analytic continuations convert the charged analytically continued solutions that appear later to their Lorentzian counterparts.


\subsubsection{Odd dimensions $D = 2n+1$}


The analytically continued Kerr--NUT metric in odd dimensions is
\bea
\ud s^2 = \sum_{\mu = 1}^n \left( \frac{X_\mu}{U_\mu}
  \mathcal{A}_\mu^2 + \frac{U_\mu}{X_\mu} \, \ud x_\mu^2 \right) - \frac{\prod_{k=1}^n a_k^2}{\prod_{\mu = 1}^n x_\mu^2} \bigg(
  \ud t - \sum_{i=1}^n \frac{\tilde{\gamma}_i}{a_i^2} \, \ud \tilde{\phi}_i \bigg) ^2 ,
\label{KNodd}
\eea
where
\bea
&&U_\mu = \sideset{}{'} \prod_{\nu = 1}^n (x_\nu^2 - x_\mu^2) , \quad
X_\mu = \frac{1}{x_\mu^2} \prod_{k=1}^n (a_k^2 - x_\mu^2) + 2 m_\mu ,
\nnr
&& \tilde{\gamma}_i = a_i^2 \prod_{\nu = 1}^n (a_i^2 - x_\nu^2) , \quad \mathcal{A}_\mu = \ud t - \sum_{i=1}^n \frac{\tilde{\gamma}_i}{a_i^2 - x_\mu^2} \, \ud \tilde{\phi}_i , \quad \tilde{\phi}_i = \frac{\phi_i}{a_i \prod^\prime{_{k=1}^n} (a_i^2 - a_k^2)} .
\label{KNodd2}
\eea
Again, $\phi_i$ are canonically normalized.  The usual Lorentzian metric, for which the negative signature arises from a vielbein that involves some $\mathcal{A}_\mu$, is recovered by the analytic continuation $x_n = \ui r$; $r$ is a radial coordinate and $m = m_n$ is a mass parameter.  Again, this will generalize to charged solutions.


\subsection{Harmonic forms}


When considering charged generalizations, various harmonic forms for the metrics (\ref{KNeven}) and (\ref{KNodd}) play prominent r\^{o}les in the solutions.  We therefore record these explicitly here, setting up the notation.  They may also help in obtaining further solutions or in other applications; for example, the 1-forms $\mathcal{A}_\mu$, which are related to harmonic 2-forms, play a key r\^{o}le in the generalized multi-Kerr--Schild form of higher-dimensional Kerr--NUT-type metrics \cite{KSharm}. 

Recall that a sufficient condition for a $p$-form $G_{(p)}$ to be harmonic is that it is closed and coclosed: $\ud G_{(p)} = 0$ and $\ud \star G_{(p)} = 0$.  First, we consider harmonic 2-forms.  For even dimensions $D = 2n$, we have potentials given by \cite{KSharm}
\ben
B_{(1)}^{(\mu)} = \frac{x_\mu}{U_\mu} \mathcal{A}_\mu ,
\een
which give rise to $n$ independent harmonic 2-forms $G_{(2)}^{(\mu)} = \ud B_{(1)}^{(\mu)}$, where each $X_\mu$ is an arbitrary function of $x_\mu$.  For odd dimensions $D = 2n+1$, we have potentials given by
\ben
B_{(1)}^{(\mu)} = \frac{1}{U_\mu} \mathcal{A}_\mu ,
\een
again giving harmonic 2-forms $G_{(2)}^{(\mu)} = \ud B_{(1)}^{(\mu)}$ for arbitrary $X_\mu$, of which $n - 1$ are independent, since $\sum_{\mu = 1}^n B_{(1)}^{(\mu)} = 0$.

We now specialize to flat space through the choice of $X_\mu$ given in (\ref{KNeven2}) and (\ref{KNodd2}) with $m_\mu = 0$ and $g = 0$.  Potentials for harmonic 1-forms are given by, for even dimensions $D = 2n$,
\ben
B^{(\mu)} = \frac{x_\mu}{U_\mu} ,
\een
and for odd dimensions $D = 2n+1$,
\ben
B^{(\mu)} = \frac{1}{U_\mu} .
\een
These give harmonic 1-forms on flat space, $G_{(1)}^{(\mu)} = \ud B^{(\mu)}$.  For even dimensions, the $B^{(\mu)}$ are all independent, giving $n$ independent harmonic 1-forms, but for odd dimensions, we have $\sum_{\mu = 1}^n B^{(\mu)} = 0$, and so these give only $n-1$ independent harmonic 1-forms.

Again in flat space, we have harmonic 2-form potentials, for even dimensions $D = 2n$,
\ben
B_{(2)}^{(\mu)} = \frac{x_\mu}{U_\mu} \, \ud t \wedge \mathcal{A}_\mu ,
\een
and for odd dimensions $D = 2n + 1$,
\ben
B_{(2)}^{(\mu)} = \frac{1}{U_\mu} \, \ud t \wedge \mathcal{A}_\mu .
\een
These give harmonic 3-forms on flat space, $H_{(3)}^{(\mu)} = \ud
B_{(2)}^{(\mu)}$.  For even dimensions, the $B_{(2)}^{(\mu)}$ are all
independent, giving $n$ independent harmonic 3-forms, but for odd
dimensions, we have $\sum_{\mu = 1}^n B_{(2)}^{(\mu)} = 0$, and so
these give only $n-1$ independent harmonic 3-forms.

It can be more convenient to instead consider a $(D-4)$-form potential for a $(D - 3)$-form field strength that is dual to a 3-form field strength.  In flat space, we could consider harmonic $(D-3)$-forms $F_{(D-3)}^{(\mu)} = \ud A_{(D-4)}^{(\mu)}$ with $H_{(3)}^{(\mu)} = \star F_{(D-3)}^{(\mu)}$.  Although it is more convenient to use Boyer--Lindquist coordinates to describe the harmonic 3-forms, Jacobi--Carter coordinates are convenient for the harmonic $(D-3)$-forms.  Instead of the time coordinate $t$ and azimuthal coordinates $\phi_i$, $i = 1, \ldots , n - 1 + \varepsilon$, we have coordinates $\psi_k$, $k = 0 , \ldots , n - 1 + \varepsilon$ defined through a linear transformation so that
\ben
\mathcal{A}_\mu = \sum_{k=0}^{n-1} A_\mu^{(k)} \, \ud \psi_k , \quad
A_\mu^{(k)} = \sum_{\substack{\nu_1 < \nu_2 < \ldots < \nu_k \\ \nu_i
    \neq \mu}} x_{\nu_1}^2 x_{\nu_2}^2 \ldots x_{\nu_k}^2 .
\label{Amu}
\een
Again in flat space, potentials for harmonic $(D-3)$-forms are given by, for even dimensions $D = 2n$,
\ben
A_{(D-4)}^{(\mu)} = \frac{\prod_{\rho = 1}^n x_\rho}{U_\mu
  x_\mu} \bigg( \sum_{\nu \neq \mu} \frac{x_\nu^2 - x_\mu^2}{x_\nu}
\, \ud x_\nu \wedge \mathcal{A}_{\mu \nu} \bigg) ^{n-2} ,
\een
and for odd dimensions $D = 2n+1$,
\ben
A_{(D-4)}^{(\mu)} = \frac{1}{U_\mu} \sum_{k=1}^n A_\mu^{(k-1)} \, \ud \psi_k \wedge \bigg( \sum_{\nu \neq \mu}
\frac{x_\nu^2 - x_\mu^2}{x_\nu} \, \ud x_\nu \wedge \mathcal{A}_{\mu \nu}
\bigg) ^{n-2} ,
\een
where
\ben
\mathcal{A}_{\mu \nu} = \sum_{k=1}^{n-1} A_{\mu \nu}^{(k-1)} \, \ud
\psi_k , \quad A_{\mu \nu}^{(k)} = \sum_{\substack{\nu_1 < \nu_2 <
    \ldots < \nu_k \\ \nu_i \neq \mu, \nu}} x_{\nu_1}^2 x_{\nu_2}^2
\ldots x_{\nu_k}^2 .
\label{Amunu}
\een
In even dimensions, we have $n$ independent $(D-3)$-forms, however in odd dimensions, we have
\ben
\sum_{\mu = 1}^n A_{(D-4)}^{( \mu )} = \ud \bigg( \frac{1}{2}
  \sum_{k=0}^n A^{(k)} \, \ud \psi_k \bigg) \wedge \ud \bigg( \sum_\mu
  x_\mu \, \ud x_\mu \wedge
\mathcal{A}_\mu \bigg) ^{n-2} \wedge \ud \psi_n ,
\een
and so these give only give $n - 1$ independent $(D - 3)$-forms.


\subsection{Equal charge solution}


We can apply the charging procedure to the higher-dimensional Kerr--NUT solution.  If both Lorentz boost parameters are set equal in the charging procedure, then the solution obtained simplifies substantially, as noted in the case without NUT parameters \cite{equalcharge}, which is the 2-charge Cveti\v{c}--Youm solution with equal charges.  In this case $s_1 = s_2 = s$ and $c_1 = c_2 = c$, and so it follows that $X = X_1 = X_2 = \ue ^{- \varphi_1 / \sqrt{2 (D-2)}}$ and $A_{(1)} = A_{(1)}^1 = A_{(1)}^2$.  There are some generalizations of these solutions to gauged supergravity; we shall recall these gauged solutions when required later.  In the Appendix, we provide the general 2-charge solution.

We present the analytically continued form of the solution using Jacobi--Boyer--Lindquist coordinates, presenting directly the metric components $g_{ab}$.  It is convenient to introduce a conformally related metric $\ud \tilde{s}^2$, which is related to the ``physical'' metric $\ud s^2$ by $\ud s^2 = H^{2/(D-2)} \, \ud \tilde{s}^2$, where $H$ is a function that we shall specify and is harmonic on flat space.


\subsubsection{Even dimensions $D = 2n$}


The analytically continued solution is
\bea
&& \ud s^2 = H^{2/(D-2)} \bigg\{ \sum_{\mu = 1}^n \bigg[ \frac{X_\mu}{U_\mu}
  \bigg( \mathcal{A}_\mu - \sum_{\nu = 1}^n \frac{2 m_\nu s^2 x_\nu}{H
      U_\nu} \mathcal{A}_\nu \bigg) ^2 + \frac{U_\mu}{X_\mu} \, \ud
  x_\mu^2 \bigg] \bigg\} , \nnr
&& X = H^{-1/(D-2)} , \quad A_{(1)} = \sum_{\mu = 1}^n \frac{2 m_\mu
  s c x_\mu}{H U_\mu} \mathcal{A}_\mu , \quad B_{(2)} = \sum_{\mu = 1}^n \frac{2
  m_\mu s^2 x_\mu}{H U_\mu} \, \ud t \wedge \sum_{i = 1}^{n - 1}
\frac{\tilde{\gamma}_i}{z_{i \mu}^2} \, \ud \tilde{\phi}_i , \nnr
\label{cKerrNUTeven}
\eea
where
\bea
&& U_\mu = \sideset{}{'} \prod_{\nu = 1}^n (x_\nu^2 - x_\mu^2) , \quad
X_\mu = - \prod_{k=1}^{n-1} (a_k^2 - x_\mu^2) + 2 m_\mu x_\mu , \quad H = 1 + \sum_{\mu = 1}^n \frac{2 m_\mu s^2 x_\mu}{U_\mu} , \nnr
&& \tilde{\gamma}_i = \prod_{\nu = 1}^n (a_i^2 -
x_\nu^2) , \quad z_{i \mu} = a_i^2 - x_\mu^2 , \quad \mathcal{A}_\mu = \ud t -  \sum_{i=1}^{n-1} \frac{\tilde{\gamma}_i}{a_i^2 - x_\mu^2} \ud
\tilde{\phi}_i .
\eea

We may simplify the solution by transforming from the time coordinate $t$ and the azimuthal $\phi_i$ coordinates to $\psi_k$ coordinates, so $\mathcal{A}_\mu$ is given by (\ref{Amu}).  However, then the expression for the 2-form potential becomes rather more complicated.  It is instead more convenient in these coordinates to give an expression for a dual $(D - 4)$-form potential (\ref{dualF}),
\ben
A_{(D - 4)} = \sum_{\mu = 1}^n \frac{2 \ui m_\mu s^2 \prod_{\rho = 1}^n x_\rho}{(n-2)! U_\mu
  x_\mu} \bigg( \sum_{\nu \neq \mu} \frac{x_\nu^2 - x_\mu^2}{x_\nu}
\, \ud x_\nu \wedge \mathcal{A}_{\mu \nu} \bigg) ^{n-2} .
\een


\subsubsection{Odd dimensions $D = 2n+1$}


The analytically continued solution is
\bea
&& \ud s^2 = H^{2/(D-2)} \bigg\{ \sum_{\mu = 1}^n \bigg[ \frac{X_\mu}{U_\mu}
  \bigg( \mathcal{A}_\mu - \sum_{\nu = 1}^n \frac{2 m_\nu s^2}{H U_\nu}
    \mathcal{A}_\nu \bigg) ^2 + \frac{U_\mu}{X_\mu} \, \ud x_\mu^2
  \bigg] \nnr
&& \qquad \qquad \qquad \quad - \frac{\prod_{k=1}^n a_k^2}{\prod_{\mu = 1}^n x_\mu^2} \bigg(
  \ud t - \sum_{i=1}^n \frac{\tilde{\gamma}_i}{a_i^2} \, \ud \tilde{\phi}_i - \sum_{\nu = 1}^n \frac{2 m_\nu s^2}{H U_\nu}
    \mathcal{A}_\nu \bigg) ^2 \bigg\} , \nnr
&& X = H^{-1/(D-2)} , \quad A_{(1)} = \sum_{\mu = 1}^n \frac{2 m_\mu
  s c}{H U_\mu} \mathcal{A}_\mu , \quad B_{(2)} = \sum_{\mu = 1}^n \frac{2 m_\mu s^2}{H U_\mu} \, \ud t \wedge \sum_{i = 1}^n \frac{\tilde{\gamma}_i}{z_{i
    \mu}^2} \, \ud \tilde{\phi}_i ,
\label{cKerrNUTodd}
\eea
where
\bea
&& U_\mu = \sideset{}{'} \prod_{\nu = 1}^n (x_\nu^2 - x_\mu^2) , \quad
X_\mu = \frac{1}{x_\mu^2} \prod_{k=1}^n (a_k^2 - x_\mu^2) + 2 m_\mu , \quad H = 1 + \sum_{\mu = 1}^n \frac{2 m_\mu s^2}{U_\mu} , \nnr
&& \tilde{\gamma}_i = a_i^2 \prod_{\nu = 1}^n (a_i^2 - x_\nu^2) , \quad z_{i \mu}^2 = a_i^2 - x_\mu^2 , \quad \mathcal{A}_\mu = \ud t - \sum_{i=1}^n \frac{\tilde{\gamma}_i}{a_i^2 - x_\mu^2} \, \ud \tilde{\phi}_i .
\eea

We may simplify the solution by transforming from the time coordinate $t$ and the azimuthal $\phi_i$ coordinates to $\psi_k$ coordinates.  The metric is then expressed as
\bea
\ud s^2 & = & H^{2/(D-2)} \bigg\{ \sum_{\mu = 1}^n \bigg[ \frac{X_\mu}{U_\mu}
  \bigg( \mathcal{A}_\mu - \sum_{\nu = 1}^n \frac{2 m_\nu s^2}{H U_\nu}
    \mathcal{A}_\nu \bigg) ^2 + \frac{U_\mu}{X_\mu} \, \ud x_\mu^2
  \bigg] \nnr
&& \qquad \qquad - \frac{\prod_{i=1}^n a_i^2}{\prod_{\mu = 1}^n
  x_\mu^2} \bigg( \sum_{k=0}^n A^{(k)} \, \ud \psi_k - \sum_{\nu = 1}^n
  \frac{2 m_\nu s^2}{H U_\nu} \mathcal{A}_\nu \bigg) ^2 \bigg\} ,
\eea
where again $\mathcal{A}_\mu$ is given by (\ref{Amu}), which also
enters the same expression for the 1-form potential, and
\ben
A^{(k)} = \sum_{\nu_1 < \nu_2 < \ldots < \nu_k} x_{\nu_1}^2
x_{\nu_2}^2 \ldots x_{\nu_k}^2 . \nnr
\een
Again, the 2-form potential becomes rather more complicated in these coordinates, and it is instead more convenient to give an expression for a dual $(D - 4)$-form potential (\ref{dualF})
\ben
A_{(D-4)} = \sum_{\mu = 1}^n \frac{2 m_\mu s^2 \prod_{i=1}^n
  a_i}{(n-2)! U_\mu} \sum_{k=1}^n A_\mu^{(k-1)} \, \ud \psi_k \wedge \bigg( \sum_{\nu \neq \mu}
\frac{x_\nu^2 - x_\mu^2}{x_\nu} \, \ud x_\nu \wedge \mathcal{A}_{\mu \nu}
\bigg) ^{n-2} .
\een


\section{Hidden symmetry}


The Kerr solution in 4 dimensions possesses, in addition to the 2 independent Killing vectors that represent time translation invariance and axisymmetry, a non-trivial rank-2 Killing--St\"{a}ckel tensor \cite{WalkerPenrose}.  This Killing tensor generates a hidden symmetry, giving rise to a constant of motion that renders geodesic motion in the spacetime integrable \cite{globalKerr}.  The solutions that we consider in this paper can all be thought of as generalizations of the Kerr solution, and we shall see that, to some extent, they possess similar properties.  We now review the key concepts relevant to hidden symmetry and apply them to the higher-dimensional charged Kerr--NUT solution that we have constructed.


\subsection{Killing tensors}


In addition to Killing vectors, the higher-dimensional Kerr--NUT--AdS metric is known to possess various higher-rank Killing tensors.  These tensors are either symmetric, known as Killing--St\"{a}ckel (or Killing or St\"{a}ckel) tensors, or antisymmetric, known as Killing--Yano (or Yano) tensors.  We focus here on the symmetric Killing--St\"{a}ckel (KS) tensors, since antisymmetric Killing--Yano tensors are not known for the charged solutions that we consider here.  It should be noted, however, that in the uncharged case of Kerr--NUT--AdS, there is a rich structure of both KS and Killing--Yano tensors, and in particular they can all be obtained from a principal conformal Killing--Yano tensor \cite{hidsymhdKNUTAdS, integgeo, KYtKt, cogmhdbh} (see also \cite{hdbhhidsymsepvar, KubiznakPhD} for a review).

A rank-2 KS tensor $K_{ab} = K_{(a b)}$ satisfies
\ben
\nabla_{(a} K_{bc)} = 0 .
\een
The metric is a trivial example of such a tensor.  More generally, a rank-2 conformal Killing--St\"{a}ckel (CKS) tensor $Q_{ab} = Q_{(a b)}$ satisfies
\ben
\nabla_{(a} Q_{bc)} = q_{(a} g_{bc)}
\een
for some covector $q_a$.  The metric multiplied by any scalar function is a
trivial example of such a tensor.  We may express $q_a$ in terms of
the CKS tensor as
\ben
q_a = \frac{1}{D+2} (\pd_a Q{^b}{_b} + 2 \nabla_b Q{^b}{_a}) .
\label{CKSvec}
\een

It is useful to introduce definitions of a rank-2 CKS tensor being of gradient type or of self-gradient type, for example as used in \cite{vsnullHJ}.  If $q_a$ is a gradient, i.e.~$q_a = \pd_a q$ for some scalar $q$, then the rank-2 CKS tensor is said to be of gradient type.  In this case, one may then construct an associated KS tensor
\ben
K_{ab} = Q_{ab} - q g_{ab} .
\label{assKS}
\een
If $q_a = Q{_a}{^b} \pd_b q$ for some scalar $q$, then the rank-2 CKS tensor is said to be of self-gradient type.  In this case, we shall later note that $Q^{ab}$ is a KS tensor for a conformally related metric.

If $k^1$ and $k^2$ are Killing vectors, then we may trivially construct a rank-2 KS tensor $K_{ab} = k{^1}{_{(a}} k{^2}{_{b)}}$.  A rank-2 KS tensor that can be expressed as a linear combination of such symmetrized products of Killing vectors and the metric is said to be reducible, otherwise it is irreducible.  Similarly, the symmetrized product of 2 conformal Killing vectors is a CKS tensor, and so one may similarly define a rank-2 CKS tensor to be reducible or irreducible.

The definitions of (C)KS tensors may be extended to tensors of rank higher than 2: a rank-$k$ KS tensor $K_{a_1 \ldots a_k} = K_{( a_1 \ldots a_k )}$ satisfies
\ben
\nabla_{( a} K_{a_1 \ldots a_k )} = 0 ,
\een
and a rank-$k$ CKS tensor $Q_{a_1 \ldots a_k} = Q_{( a_1 \ldots a_k )}$ satisfies an equation of the form
\ben
\nabla_{( a} Q_{a_1 \ldots a_k )} = q_{( a a_1 \ldots a_{k-2}} g_{a_{k-1} a_k )} .
\een

With every Killing vector $k^a$, there is a quantity conserved along geodesic motion, $k^a P_a$, where $P_a$ is the canonical momentum.  Similarly, a KS tensor $K^{a_1 \ldots a_k}$ gives rise to a quantity conserved along geodesic motion, $K^{a_1 \ldots a_k} P_{a_1} \ldots P_{a_k}$.  A CKS tensor $Q^{a_1 \ldots a_k}$ also gives rise to a conserved quantity, but only along null geodesics, in which case $Q^{a_1 \ldots a_k} P_{a_1} \ldots P_{a_k}$ is conserved.

It is well-known that if $k^a$ is a conformal Killing vector for the metric $\ud s^2$, then $k^a$ is a conformal Killing vector for any conformally related metric, $\ud \tilde{s}^2 = \ue ^{2 \Omega} \ud s^2$.  Similarly, if $Q^{ab}$ is a rank-2 CKS tensor for the metric $\ud s^2$, then $Q^{ab}$ is a rank-2 CKS tensor for the metric $\ud \tilde{s}^2 = \ue ^{2 \Omega} \ud s^2$.  It is straightforward to generalize the result to CKS tensors of any rank: if $Q^{a_1 \ldots a_k}$ is a CKS tensor for the metric $\ud s^2$, satisfying
\ben
\nabla^{(a_1} Q^{a_2 \ldots a_{k+1})} = q^{(a_1 \ldots a_{k-1}} g^{a_k a_{k+1})} ,
\een
then $Q^{a_1 \ldots a_k}$ is a CKS tensor for the metric $\ud \tilde{s}^2 = \ue ^{2 \Omega} \ud s^2$ , satisfying
\ben
\tilde{\nabla}^{(a_1} Q^{a_2 \ldots a_{k+1})} = \tilde{q}^{(a_1 \ldots
  a_{k-1}} g^{a_k a_{k+1})} , \quad \tilde{q}^{a_1 \ldots a_{k-1}} =
q^{a_1 \ldots a_{k-1}} + k Q^{a_1 \ldots a_{k-1} b} \pd_b \Omega .
\een
Specializing to the rank-2 case again, we see that $Q^{ab}$ being a CKS tensor of self-gradient type for some metric $\ud s^2$, with $q^a = Q^{ab} \pd_b q$, is equivalent to $Q^{ab}$ being a KS tensor for the conformally related metric $\ud \tilde{s}^2 = \ue^{-q} \ud s^2$.


\subsection{Integrability and separability structure}


The Schouten--Nijenhuis (SN) bracket of 2 symmetric contravariant tensors $A^{a_1 \ldots a_m} = A^{(a_1 \ldots a_m)}$ and $B^{a_1 \ldots a_n} = B^{(a_1 \ldots a_n)}$, respectively of ranks $m$ and $n$, is defined as
\bea
[A, B]{_{\textrm{SN}}}{^{a_1 \ldots a_{m+n-1}}} & = & m A^{b ( a_1 \ldots
  a_{m-1}} \nabla_b B^{a_m \ldots a_{m+n-1} )} - n B^{b ( a_1 \ldots
  a_{m-1}} \nabla_b A^{a_m \ldots a_{m+n-1} )} \nnr
& = & m A^{b ( a_1 \ldots
  a_{m-1}} \pd_b B^{a_m \ldots a_{m+n-1} )} - n B^{b ( a_1 \ldots
  a_{m-1}} \pd_b A^{a_m \ldots a_{m+n-1} )} .
\eea
Under this bracket, contravariant symmetric tensors form a graded Lie algebra.  If $A$ is simply a vector, then the SN bracket reduces to the Lie derivative of $B$ along $A$.  Just as we may define a Killing vector $k$ by the vanishing of the Lie derivative of the metric along $k$, i.e.~$\mathcal{L}_k g_{ab} = 0$, we may alternatively define a KS tensor $K$ by the vanishing of the SN bracket of $K$ and the metric, i.e.~$[K, g]_{\textrm{SN}} = 0$, with similar definitions for their conformal counterparts.

If a metric possesses $D - r$ Killing vectors and $r - 1$ irreducible KS tensors, then, together with the metric, there are $D$ independent constants of motion, rendering geodesic motion integrable.  If all of the SN brackets of these tensors vanish, then geodesic motion is completely integrable in the sense of Liouville; equivalently, the conserved quantities associated with these tensors Poisson commute, i.e.~are in involution.

If a metric possesses enough Killing vectors and KS tensors that satisfy certain conditions, then the metric possesses a separability structure, which implies that the HJE for geodesic motion is separable \cite{sepstruc, sepHJgr, KtHJeqns, sepvarHJeqn}.  Furthermore, for an Einstein metric, a separability structure also implies that the KGE separates.  If there are $D - r$ independent Killing vectors and $r - 1$ independent irreducible rank-2 KS tensors with all SN brackets vanishing, and if also the irreducible KS tensors and the metric, when viewed as linear maps, have $r$ orthogonal eigenvectors that commute with each other and with the Killing vectors, then there is a separability structure $\mathcal{S}_{D - r}$.

We shall consider charged solutions for which irreducible CKS tensors, but not irreducible KS tensors, are known.  One might then expect that the HJE
\ben
\mathcal{H} \left( x^a , \frac{\pd W}{\pd x^a} \right) = - \frac{1}{2}
\mu^2,
\een
where $W$ is Hamilton's characteristic function and $\mu$ represents a particle mass, separates for only null geodesics, i.e.~for the null HJE, which has $\mu = 0$.  Although separability structures for the HJE were studied a long time ago, only more recently has there been study of the null HJE \cite{vsnullHJ}, where criteria for separability are given.  One such criterion is that the null HJE is separable in the coordinates $x^a$ if and only if there is a nowhere-vanishing function $\Lambda (x^a, p_a)$ such that the HJE for the modified Hamiltonian $\tilde{\mathcal{H}} = \mathcal{H} / \Lambda$ is separable.  For our study of geodesics, the Hamiltonian is
\ben
\mathcal{H} = \frac{1}{2} g^{ab} \frac{\pd W}{\pd x^a} \frac{\pd
  W}{\pd x^b} .
\een
If the HJE separates for some conformally related metric $\tilde{g}_{ab}$, then the null HJE for the original ``physical'' metric $g_{ab}$ separates.


\subsection{Charged Kerr--NUT symmetries}


We consider here the charged Kerr--NUT solution with 2 equal charges.  Its analytically continued metric is given in (\ref{cKerrNUTeven}) for even dimensions and in (\ref{cKerrNUTodd}) for odd dimensions.  Setting $s = 0$ in the charged solution recovers the higher-dimensional Kerr--NUT metric \cite{genKNUTAdSalld}, and so we first recall some facts about its metric given in analytically continued form by (\ref{KNeven}) for even dimensions and (\ref{KNodd}) for odd dimensions.  In fact, we can consider the wider family of metrics for which each function $X_\mu$ that appears in the metric is an arbitrary function of $x_\mu$.  Therefore we include, for example, higher-dimensional Kerr--NUT--AdS, but not the charged solutions of this paper, for which there are various extra factors of harmonic functions.  It is clear that, in $D = 2n + \varepsilon$ dimensions, there are $n + \varepsilon$ Killing vectors given by $\pd / \pd \psi_k$, $k = 0, \ldots, n - 1 + \varepsilon$.  The family of metrics also possesses $n-1$ irreducible rank-2 KS tensors, found in \cite{integgeo, KYtKt}; these may be concisely expressed in terms of a simple set of vielbeins.  The metric is trivially also a KS tensor.  The conditions for a separability structure are satisfied, and one can check directly that the HJE and KGE separate \cite{sepHJKGKNUTAdS}.

For investigating whether or not there are (C)KS tensors for the charged solution that we have constructed, we restrict ourselves to the simpler equal charge case, which can be presented with vielbeins that generalize those of the uncharged case in a simple way.  The (analytically continued) ``physical'' metric is
\ben
\ud s^2 = \sum_{\mu = 1}^n (e^\mu e^\mu + e^{\hat{\mu}} e^{\hat{\mu}})
- \varepsilon e^{\hat{0}} e^{\hat{0}} ,
\een
where the vielbeins are
\bea
&& e^\mu = H^{1/(D-2)} \sqrt{\frac{U_\mu}{X_\mu}} \, \ud x_\mu , \quad
e^{\hat{\mu}} = H^{1/(D-2)} \sqrt{\frac{X_\mu}{U_\mu}} \bigg(
  \mathcal{A}_\mu - \sum_{\nu = 1}^n \frac{2 N_\nu s^2}{H U_\nu}
  \mathcal{A}_\nu \bigg) , \nnr
&& e^{\hat{0}} = H^{1/(D-2)} \frac{c}{P} \bigg( \sum_{k=0}^n A^{(k)} \, \ud \psi_k - \sum_{\nu =
    1}^n \frac{2 N_\nu s^2}{H U_\nu} \mathcal{A}_\nu \bigg) .
\eea
We have defined $N_\mu = m_\mu x_\mu^{1 - \varepsilon}$, so $N_\mu = m_\mu x_\mu$ for even dimensions, and $N_\mu = m_\mu$ for odd dimensions.  We have also, in odd dimensions, $c = \prod_{i=1}^n a_i$, $P = \prod_{\rho = 1}^n x_\rho$; note that $P^2 = A^{(n)}$.

However, from the geometrical viewpoint of hidden symmetries, it seems that the metric $\ud \tilde{s}^2 = H^{-2/(D-2)} \, \ud s^2$, conformally related to the ``physical'' metric $\ud s^2$, is more fundamental, in that it possesses irreducible rank-2 KS tensors that generalize those of the higher-dimensional Kerr--NUT metric.  The conformally related metric is
\ben
\ud \tilde{s}^2 = \sum_{\mu = 1}^n (\tilde{e}^\mu \tilde{e}^\mu +
\tilde{e}^{\hat{\mu}} \tilde{e}^{\hat{\mu}}) - \varepsilon
\tilde{e}^{\hat{0}} \tilde{e}^{\hat{0}} ,
\een
where the vielbeins are
\bea
&& \tilde{e}^\mu = \sqrt{\frac{U_\mu}{X_\mu}} \, \ud x_\mu , \quad
\tilde{e}^{\hat{\mu}} = \sqrt{\frac{X_\mu}{U_\mu}} \bigg(
  \mathcal{A}_\mu - \sum_{\nu = 1}^n \frac{2 N_\nu s^2}{H U_\nu}
  \mathcal{A}_\nu \bigg) , \nnr
&& \tilde{e}^{\hat{0}} = \frac{c}{P} \bigg( \sum_{k=0}^n A^{(k)} \, \ud \psi_k - \sum_{\nu = 1}^n
  \frac{2 N_\nu s^2}{H U_\nu} \mathcal{A}_\nu \bigg) .
\eea
Denoting tangent space indices by $A = \{ \mu, \hat{\mu} \}$ in even dimensions and $A = \{ \mu, \hat{\mu}, \hat{0} \}$ in odd dimensions, we have $\tilde{e}^A = H^{-1/(D-2)} e^A$.

Henceforth, we shall make use of these 2 types of metric.  We shall refer to the Einstein frame metric $\ud s^2$ (the ``physical'' metric) as one that solves the usual supergravity field equations, derived from a Lagrangian of the form $\mathcal{L}_D = R \star 1 + \ldots$, including any analytic continuations that may be convenient.  We shall refer to the Jordan frame metric $\ud \tilde{s}^2$ as one that is related to the Einstein frame metric by $\ud \tilde{s}^2 = H^{-2/(D-2)} \, \ud s^2$, or with some similar conformal factor, again including any analytic continuations.  In the context of non-critical string theory in arbitrary dimension $D$, $\ud \tilde{s}^2$ would be known as the string-frame metric.

From the first Cartan structure equation $\ud \tilde{e}^A + \tilde{\omega}{^A}{_B} \wedge \tilde{e}^B = 0$ and the antisymmetry $\omega_{AB} = - \omega_{BA}$, the connection 1-forms are
\bea
\tilde{\omega}_{\mu \nu} & = & (1 - \delta_{\mu \nu}) \left( -
  \sqrt{\frac{X_\nu}{U_\nu}} \frac{x_\nu}{x_\mu^2 - x_\nu^2}
  \tilde{e}^\mu - \sqrt{\frac{X_\mu}{U_\mu}} \frac{x_\mu}{x_\mu^2 -
    x_\nu^2} \tilde{e}^\nu \right) , \nnr
\tilde{\omega}_{\mu \hat{\nu}} & = & \delta_{\mu \nu} \left[ - H
  \pd_\mu \left( \frac{1}{H} \sqrt{\frac{X_\mu}{U_\mu}} \right)
  \tilde{e}^{\hat{\mu}} + \frac{1}{2} \sum_{\rho \neq \mu}
  \sqrt{\frac{X_\mu}{U_\mu}} \pd_\mu [\log (H U_\rho)]
  \, \tilde{e}^{\hat{\rho}} - \varepsilon \frac{c}{P} \left( \frac{1}{x_\mu} + \frac{1}{2} \pd_\mu \log H
\right) \tilde{e}^{\hat{0}} \right] \nnr
&& + (1 - \delta_{\mu \nu}) \left( - \sqrt{\frac{X_\mu}{U_\mu}}
  \frac{x_\mu}{x_\mu^2 - x_\nu^2} \tilde{e}^{\hat{\nu}} + \frac{1}{2}
  \sqrt{\frac{X_\nu}{U_\nu}} \pd_\mu [\log (H U_\nu)]
  \, \tilde{e}^{\hat{\mu}} \right) \nnr
\tilde{\omega}_{\hat{\mu} \hat{\nu}} & = & (1 - \delta_{\mu \nu})
\left( - \frac{1}{2} \sqrt{\frac{X_\nu}{U_\nu}} \pd_\mu [\log (H
  U_\nu)] \, \tilde{e}^\mu + \frac{1}{2} \sqrt{\frac{X_\mu}{U_\mu}}
  \pd_\nu [\log (H U_\mu)] \, \tilde{e}^\nu \right) , \nnr
\tilde{\omega}_{\mu \hat{0}} & = & - \frac{c}{P} \left( \frac{1}{x_\mu} + \frac{1}{2}
  \pd_\mu \log H \right) \tilde{e}^{\hat{\mu}} +
\sqrt{\frac{X_\mu}{U_\mu}} \frac{1}{x_\mu} \tilde{e}^{\hat{0}} , \quad \tilde{\omega}_{\hat{\mu} \hat{0}} = \frac{c}{P} \left( \frac{1}{x_\mu} + \frac{1}{2}
  \pd_\mu \log H \right) \tilde{e}^\mu ,
\eea
where the last 2 sets of 1-forms, $\tilde{\omega}_{\mu \hat{0}}$ and $\tilde{\omega}_{\hat{\mu} \hat{0}}$, exist for odd dimensions only.  We may then compute covariant derivatives with ease through the Ricci rotation coefficients $\tilde{\omega}_{ABC}$, defined by $\tilde{\omega}{^A}{_B} = \tilde{\omega}{^A}{_{BC}} \tilde{e}^C$.  Note that these expressions for the coefficients do not depend much on the details of the functions $X_\mu$, which is why the discussion of symmetries here is valid for $X_\mu$ being arbitrary functions of $x_\mu$.  In the uncharged case $H = 1$, these connection 1-forms reduce to those of \cite{KNdScurv}.

Again, there are $n + \varepsilon$ Killing vectors given by $\pd / \pd \psi_k$, $k = 0 , \ldots, n - 1 + \varepsilon$.  The Jordan frame metric $\ud \tilde{s}^2$ also has $n-1$ irreducible KS tensors given by
\ben
\widetilde{K}^{(j)} = \sum_{\mu = 1}^n A_\mu^{(j)} (\tilde{e}^\mu
\tilde{e}^\mu + \tilde{e}^{\hat{\mu}} \tilde{e}^{\hat{\mu}}) -
\varepsilon A^{(j)} \tilde{e}^{\hat{0}} \tilde{e}^{\hat{0}} ,
\een
where $j = 1 , \ldots , n-1$.  The metric, which is trivially a KS tensor, is obtained from $j = 0$.  Again, each function $X_\mu$ that appears in the metric can be an arbitrary function of $x_\mu$.  As a consequence, the Einstein frame metric has $n-1$ irreducible CKS tensors given by
\ben
Q^{(j)} = H^{2/(D-2)} \bigg( \sum_{\mu = 1}^n A_\mu^{(j)} (e^\mu e^\mu
  + e^{\hat{\mu}} e^{\hat{\mu}}) - \varepsilon A^{(j)} e^{\hat{0}}
  e^{\hat{0}} \bigg) .
\een
The associated covector has non-vanishing components
\ben
q{^{(j)}}{_\mu} = \frac{2}{D-2} H^{(4-D)/(D-2)}
\pd_\mu ( A_\mu^{(j)} H ) .
\een
For charged solutions, which have $H \neq 1$, then only in the $D = 4$ case, for which we must have $j = 1$, is $q{^{(j)}}{_a}$ a gradient.  In this case, we have $q{^{(1)}}{_a} = \pd_a q^{(1)}$, with
\ben
q^{(1)} = \frac{2 m_1 s^2 x_1 x_2^2}{x_2^2 - x_1^2} + \frac{2 m_2 s^2
  x_2 x_1^2}{x_1^2 - x_2^2} ,
\een
and so we can obtain an irreducible KS tensor for the Einstein frame metric,
\ben
K = Q^{(1)} - q^{(1)} \, \ud s^2 = x_2 (x_2 + 2 m_2 s^2) (e^1 e^1 +
e^{\hat{1}} e^{\hat{1}}) + x_1 (x_1 + 2 m_1 s^2) (e^2 e^2 +
e^{\hat{2}} e^{\hat{2}}) .
\een

We find that the SN brackets amongst Killing vectors and KS tensors of the Jordan frame metric $\ud \tilde{s}^2$ vanish.  $\pd / \pd x_\mu$ are clearly orthogonal eigenvectors of the CKS tensors, when viewed as linear maps, and commute with each other and with the Killing vectors.  Therefore we have a separability structure, and so the HJE for the Jordan frame metric $\ud \tilde{s}^2$ is separable, and the HJE for null geodesics of the Einstein frame metric $\ud s^2$ is separable.  The SN bracket may be expressed in terms of partial derivatives, and so it follows that the SN brackets amongst Killing vectors and the induced (C)KS tensors of the Einstein frame metric also vanish.

In 4 dimensions, without any NUT parameter, the 2-charge Cveti\v{c}--Youm solution with both charges equal was originally obtained in \cite{rotcbhhetst}.  A rank-2 KS tensor for the Einstein frame metric has previously been obtained in \cite{HiokiMiyamoto}, in which the separability structure conditions were verified and the Einstein frame HJE was directly separated.


\subsection{Charged Kerr--NUT separability}


We have seen that the charged Kerr--NUT solution with 2 equal charges has an Einstein frame metric that is conformally related to a Jordan frame metric that possesses a separability structure.  Therefore the HJE for the Jordan frame metric is separable, and so the null HJE for the Einstein frame metric is separable.  For an Einstein metric, the existence of a separability structure also implies the separability of the KGE.  Since we are including charge in our solutions, the separability structure does not guarantee such separation, but we shall see that there is separation for the massless KGE of the Einstein frame metric.  We now proceed with a direct demonstration of the separability of these equations.

We first need to compute inverse metrics, which is a straightforward task for our choice of coordinates and vielbeins.  For the Jordan frame metric $\ud \tilde{s}^2 = H^{-2/(D-2)} \, \ud s^2$, the inverse metric is
\bea
\left( \frac{\pd}{\pd \tilde{s}} \right) ^2 & = & \sum_{\mu = 1}^n
\bigg[ \frac{X_\mu}{U_\mu} \left( \frac{\pd}{\pd x_\mu} \right) ^2 +
  \frac{1}{X_\mu U_\mu} \bigg( \sum_{k=0}^{n-1+\varepsilon}
    (-x_\mu^2)^{n-1-k} \frac{\pd}{\pd \psi_k} + 2 N_\mu s^2
    \frac{\pd}{\pd \psi_0} \bigg) ^2 \bigg] \nnr
&& - \frac{\varepsilon}{c^2 P^2} \left( \frac{\pd}{\pd \psi_n} \right) ^2 .
\eea
The inverse of the Einstein frame metric $\ud s^2$ is $(\pd / \pd s)^2 = H^{-2/(D-2)} (\pd / \pd \tilde{s})^2$.


\subsubsection{Hamilton--Jacobi equation}


The HJE for geodesic motion on the Jordan frame metric is
\ben
\frac{\pd S}{\pd \lambda} + \frac{1}{2} \tilde{g}^{ab} \, \pd_a S \, \pd_b S = 0 ,
\label{HJeqn}
\een
where $\lambda$ is an affine parameter.  To demonstrate the separation directly, we essentially follow the calculation of \cite{sepHJKGKNUTAdS} for the higher-dimensional Kerr--NUT--AdS metric, so we shall be slightly brief here.  We consider a separable solution of the HJE,
\ben
S = \frac{1}{2} \mu^2 \lambda + \sum_{\mu = 1}^n S_\mu (x_\mu) +
\sum_{k=0}^{n-1+\epsilon} \Psi_k \psi_k ,
\label{HJsep}
\een
where $\Psi_k$ are constants that arise from momenta conjugate to the ignorable coordinates $\psi_k$.  Substituting the separable solution into the HJE gives
\ben
\sum_{\mu = 1}^n \frac{F_\mu (x_\mu)}{U_\mu} = - \mu^2 + \frac{\varepsilon
  \Psi_n^2}{c^2 P^2} ,
\label{HJsubs}
\een
where
\ben
F_\mu = X_\mu (S'_\mu)^2 + \frac{1}{X_\mu} \left( \sum_{k=0}^{n - 1 +
    \varepsilon} (-x_\mu^2)^{n-1-k} \Psi_k + 2 N_\mu s^2 \Psi_0
\right) ^2 ,
\een
and $S'_\mu$ denotes the derivative of $S_\mu$ with respect to $x_\mu$.  It can be shown \cite{sepHJKGKNUTAdS} that (\ref{HJsubs}) is satisfied by
\ben
F_\mu = \sum_{k=0}^{n-1+\varepsilon} c_k (-x_\mu^2)^{n-1-k} ,
\een
where $c_k$ are arbitrary constants, with $c_0 = - \mu^2$ and, in odd dimensions, $c_n = - \Psi_n^2 / c^2$.  We therefore have
\ben
(S'_\mu)^2 = - \frac{1}{X_\mu^2} \left( \sum_{k=0}^{n - 1 +
    \varepsilon} (-x_\mu^2)^{n-1-k} \Psi_k + 2 N_\mu s^2 \Psi_0
\right) ^2 + \frac{1}{X_\mu} \sum_{k=0}^{n-1+\varepsilon} c_k
(-x_\mu^2)^{n-1-k} ,
\een
which can be solved by quadratures, and so we have demonstrated the separation.

We now return to the HJE for the Einstein frame metric, with $\tilde{g}^{ab}$ in (\ref{HJeqn}) replaced by $g^{ab} = H^{-2/(D-2)} \tilde{g}^{ab}$.  Again, we substitute the separable solution (\ref{HJsep}) into the HJE to give, instead of (\ref{HJsubs}),
\ben
\sum_{\mu = 1}^n \frac{F_\mu}{U_\mu} = - H^{2/(D-2)} \mu^2 +
\frac{\varepsilon \Psi_n^2}{c^2 P^2} .
\een
For the null HJE, which has $\mu = 0$, separation proceeds in the same manner as for the Jordan frame metric $\ud \tilde{s}^2$.  For $\mu \neq 0$ and $D \geq 5$, no choice of $F_\mu$ can satisfy this equation, because $H^{2/(D-2)}$ is not a rational function.  For $\mu \neq 0$ and $D = 4$, we have
\ben
F_\mu = - \mu^2 ( - x_\mu^2 + 2 m_\mu s^2 x_\mu) + c_1 ,
\een
and so
\ben
(S'_\mu)^2 = - \frac{1}{X_\mu^2} [(-x_\mu^2 + 2 m_\mu s^2 x_\mu)\Psi_0
+ \Psi_1]^2 + \frac{1}{X_\mu} [ - \mu^2 (-x_\mu^2 + 2 m_\mu s^2 x_\mu) +
c_1 ] , 
\een
which again can be solved by quadratures.


\subsubsection{Klein--Gordon equation}


The (minimally coupled) KGE for the Einstein frame metric is
\ben
\square \Phi = \frac{1}{\sqrt{|g|}} \pd_a (\sqrt{|g|} g^{ab} \pd_b \Phi) = \mu^2 \Phi , 
\een
where $g = \det(g_{ab})$.  For this metric, we have
\ben
\det (g_{ab}) = H^{4/(D-2)} U^2 ( c P ) ^{2 \varepsilon} , \quad U = \prod_{\mu < \nu} (x_\mu^2 - x_\nu^2) .
\een
The direct demonstration of separability again follows \cite{sepHJKGKNUTAdS} closely, so we shall be brief here.  We consider a separable solution of the KGE,
\ben
\Phi = \prod_{\mu = 1}^n R_\mu (x_\mu) \prod_{k=0}^{n-1+\varepsilon}
\ue^{\ui \Psi_k \psi_k} ,
\een
where again $\Psi_k$ are constants that arise from momenta conjugate to the ignorable coordinates $\psi_k$.  Substituting the separable solution into the KGE gives
\ben
\sum_{\mu = 1}^n \frac{G_\mu (x_\mu)}{U_\mu} = H^{2/(D-2)} \mu^2 ,
\label{KGsubs}
\een
where
\bea
G_\mu & = & \frac{(X_\mu R'_\mu)'}{R_\mu} + \frac{\varepsilon
    X_\mu R'_\mu}{x_\mu R_\mu} -
\frac{1}{X_\mu} \left( \sum_{k=0}^{n-1+\varepsilon}
  (-x_\mu^2)^{n-1-k} \Psi_k + 2 N_\mu s^2 \Psi_0 \right) ^2 +
\frac{\varepsilon \Psi_n^2}{c^2 x_\mu^2} .
\eea
For $\mu = 0$, the massless KGE, separation is possible.  It can be shown \cite{sepHJKGKNUTAdS} that (\ref{KGsubs}) is satisfied for $\mu = 0$ by 
\ben
G_\mu = \sum_{k=1}^{n-1} b_k (-x_\mu^2)^{n-1-k} ,
\een
where $b_k$ are arbitrary constants, with $b_n = \Psi_n^2 / c^2$ in odd dimensions.  We therefore have
\bea
&& (X_\mu R'_\mu)' + \frac{\varepsilon X_\mu R'_\mu}{x_\mu} -
\frac{R_\mu}{X_\mu} \left( \sum_{k=0}^{n-1+\varepsilon}
  (-x_\mu^2)^{n-1-k} \Psi_k + 2 N_\mu s^2 \Psi_0 \right) ^2 - \sum_{k=1}^{n-1+\varepsilon} b_k (-x_\mu^2)^{n-1-k} R_\mu = 0 . \nnr
\eea
These are second order ordinary differential equations for $R_\mu (x_\mu)$, and so we have demonstrated the separation.  For $\mu \neq 0$ and $D \geq 5$, no choice of $G_\mu$ can satisfy (\ref{KGsubs}), because $H^{2/(D-2)}$ is not a rational function.  For $\mu \neq 0$ and $D = 4$, we have
\ben
G_\mu = \mu^2 (- x_\mu^2 + 2 m_\mu s^2 x_\mu) + b_1 ,
\een
and so
\ben
(X_\mu R'_\mu)' -
\frac{R_\mu}{X_\mu} [(-x_\mu^2 + 2 m_\mu s^2 x_\mu) \Psi_0 + \Psi_1]
^2 - [\mu^2 (- x_\mu^2 + 2 m_\mu s^2 x_\mu) + b_1] R_\mu = 0 ,
\een
which demonstrates the separation.


\section{Equal charge gauged supergravity black holes}


We now extend this study of Killing tensors to various charged and rotating black hole solutions in gauged supergravity that have some combination of charges equal, generalizing those of ungauged supergravity.  We henceforth express their $D$-dimensional Lorentzian metrics in terms of vielbeins as
\ben
\ud s^2 = - e^0 e^0 + \sum_{A = 1}^{D - 1} e^A e^A . 
\een
(C)KS tensors for some of these black holes have appeared in the literature already \cite{intcrotbh45, sepmcbh, symKerrAdS5}; here, we obtain further (C)KS tensors and present those that correspond to equal charge black holes in a more unified manner, and see that they have concise expressions when expressed using an appropriate choice of vielbeins.  Furthermore, we see that these are induced from KS tensors of a conformally related Jordan frame metric.

Again, in all of these cases, there is no dependence on the precise expressions for the functions $X_\mu$ that appear in the metric as far as obtaining Killing tensors is concerned, provided that each $X_\mu$ is a function of $x_\mu$.  We again find that for the Jordan frame metric geodesic motion is completely integrable in the sense of Liouville and furthermore the conditions for a separability structure are satisfied.  Therefore the HJE of the Jordan frame metric and hence also the null HJE of the Einstein frame metric separate.  Separability of the HJE has previously been directly verified in \cite{intcrotbh45} for the $D = 4$ example considered here, and in \cite{sepmcbh, symKerrAdS5} for some of the $D = 5$ examples considered here.


\subsection{Four dimensions}



\paragraph{Kerr--NUT--AdS solution with pairwise equal charges:}


In $D = 4$, there is a $\uU (1)^4$ gauged supergravity theory, and a black hole solution with pairwise equal charges was found in \cite{crotbh4d}.  In this pairwise equal charge case with both a mass parameter and a NUT parameter, an irreducible KS tensor was found in \cite{intcrotbh45}, and agrees with that presented here, up to a reducible KS tensor\footnote{Note that in \cite{intcrotbh45} the expression for $K^{tt}$ should contain $(u_1 u_2)^2$ instead of $u_1 u_2$.}.  The Lorentzian metric has
\bea
e^0 & = & H^{-1/2} \sqrt{\frac{R}{r^2 + y^2}} (\ud t' + y_1 y_2
\, \ud \psi) , \quad e^1 = H^{1/2} \sqrt{\frac{r^2 + y^2}{R}}
\, \ud r , \nnr
e^2 & = & H^{-1/2} \sqrt{\frac{Y}{r^2 + y^2}} (\ud t' - r_1 r_2
\, \ud \psi) , \quad e^3 = H^{1/2} \sqrt{\frac{r^2 + y^2}{Y}}
\, \ud y ,
\eea
and
\bea
&& R = a^2 + r^2 - 2 m r + g^2 r_1 r_2 (r_1 r_2 + a^2) , \quad Y = a^2
- y^2 + 2 \ell y + g^2 y_1 y_2 (y_1 y_2 - a^2) , \nnr
&& r_I = r + 2 m s_I^2 , \quad y_I = y + 2 \ell s_I^2 , \quad H =
\frac{r_1 r_2 + y_1 y_2}{r^2 + y^2} .
\eea
The Jordan frame metric, which has vielbeins $\tilde{e}^A = H^{-1/2} e^A$, has an irreducible KS tensor
\ben
\widetilde{K} = y^2 (- \tilde{e}^0 \tilde{e}^0 + \tilde{e}^1 \tilde{e}^1)
- r^2 (\tilde{e}^2 \tilde{e}^2 + \tilde{e}^3 \tilde{e}^3) .
\een
Returning to the Einstein frame metric, we therefore have an irreducible CKS tensor
\ben
Q = H [ y^2 ( - e^0 e^0 + e^1 e^1 ) - r^2 (e^2 e^2 + e^3 e^3) ] .
\een
This is of gradient type, since $q_a = \pd_a q$ with
\ben
q = \frac{r_1 r_2 y^2 - y_1 y_2 r^2}{r^2 + y^2} ,
\een
and so we have an associated irreducible KS tensor
\ben
K = y_1 y_2 (- e^0 e^0 + e^1 e^1) - r_1 r_2 (e^2 e^2 + e^3 e^3) .
\een


\paragraph{Pleba\'{n}ski--Demia\'{n}ski solution with pairwise equal charges:}


A general solution that includes the Kerr--NUT--AdS solution and the C-metric is the Pleba\'{n}ski--Demia\'{n}ski solution \cite{classsolEMeq, PlebDem}.  There is correspondingly a generalization, to include an acceleration parameter, of the Kerr--NUT solution with pairwise equal charges in the $\uU (1)^4$ ungauged supergravity theory; no gauged generalization is currently known.  The Lorentzian metric has
\bea
&& e^0 = \frac{H^{-1/2}}{1 - ry} \sqrt{\frac{R}{r^2 + y^2}} (\ud t' +
y_1 y_2 \, \ud \psi) , \quad e^1 = \frac{H^{1/2}}{1 - ry}
\sqrt{\frac{r^2 + y^2}{R}} \, \ud r , \nnr
&& e^2 = \frac{H^{-1/2}}{1 - ry} \sqrt{\frac{Y}{r^2 + y^2}} (\ud t' -
r_1 r_2 \, \ud \psi) , \quad e^3 = \frac{H^{1/2}}{1 - ry}
\sqrt{\frac{r^2 + y^2}{Y}} \, \ud y ,
\eea
and
\bea
&& R = \gamma - 2 m r + \epsilon r^2 - 2 \ell r^3 - \gamma r^4 , \quad Y = \gamma + 2 \ell y - \epsilon y^2 + 2 m y^3 - \gamma y^4 , \nnr
&& r_I = r + 2 m s_I^2 , \quad y_I = y + 2 \ell s_I^2 , \quad H =
\frac{r_1 r_2 + y_1 y_2}{r^2 + y^2} .
\eea

Considering the general form of the vielbeins compared to those of the solution without the acceleration parameter, we see that they differ by an overall conformal factor; less importantly for our purposes, the functions $R$ and $Y$ also differ.  As a consequence, we can immediately write down (C)KS tensors for this solution with an acceleration parameter.  Therefore, for the Jordan frame metric, which has vielbeins $\te ^A = H^{-1/2} (1 - ry) e^A$, there is an irreducible KS tensor
\ben
\widetilde{K} = y^2 (- \te ^0 \te ^0 + \te ^1 \te ^1) - r^2 (\te ^2 \te ^2 + \te ^3 \te ^3) .
\een
Returning to the Einstein frame metric, we therefore have an irreducible CKS tensor
\ben
Q = H (1 - ry)^{-2} [ y^2 ( - e^0 e^0 + e^1 e^2 ) - r^2 (e^2 e^2 + e^3 e^3) ] ,
\een
which is not of gradient type.


\subsection{Five dimensions}


In $D = 5$, there is a $\uU (1)^3$ gauged supergravity theory, and various black hole solutions are known.  The ungauged supergravity black hole with equal charges that we have considered previously, namely the 2-charge Cveti\v{c}--Youm solution with equal charges, should be regarded as having 3 charge parameters with $\delta_1 = \delta_2$ and $\delta_3 = 0$.  We consider here gauged solutions with charge parameters such that $\delta_1 = \delta_2$ and $\delta_3 = 0$, which possesses an irreducible CKS tensor, and with $\delta_1 = \delta_2 = \delta_3$, which is the only gauged solution known to possess an irreducible KS tensor.  There is a further generalization \cite{newrotbh5d} with $\delta_1 = \delta_2$ and $\delta_3$ arbitrary that interpolates between these 2 solutions.


\paragraph{Two equal charges, third charge zero, $\delta_1 = \delta_2$, $\delta_3 = 0$:}


The 5-dimensional 2-charge Cveti\v{c}--Youm solution with both charges equal has a generalization to gauged supergravity \cite{5dgsugrabhind}.  An irreducible CKS tensor was found in \cite{sepmcbh}, and agrees with that presented here, up to a reducible KS tensor.  The Lorentzian metric has
\bea
&& e^0 = H^{-2/3} \sqrt{\frac{R}{r^2 + y^2}} \mathcal{A} , \quad e^1 =
H^{1/3} \sqrt{\frac{r^2 + y^2}{R}} \, \ud r , \nnr
&& e^2 = H^{1/3} \sqrt{\frac{Y}{r^2 + y^2}} \left( \ud t' - r^2
 \, \ud \psi_1 - \frac{q}{H (r^2 + y^2)} \mathcal{A} \right) ,
\quad e^3 = H^{1/3} \sqrt{\frac{r^2 + y^2}{Y}} \, \ud y , \nnr
&& e^4 = H^{1/3} \frac{a b}{r y} \left( \ud t' + (y^2 - r^2)
\,  \ud \psi_1 - r^2 y^2 \, \ud \psi_2 - \frac{q}{H (r^2 +
    y^2)} \mathcal{A} \right) ,
\eea
and
\bea
&& R = \frac{(1 + g^2 r^2) (r^2 + a^2) (r^2 + b^2)}{r^2} + q g^2 (2
r^2 + a^2 + b^2) + q^2 g^2 - 2m , \nnr
&& Y = - \frac{(1 - g^2 y^2) (a^2 - y^2) (b^2 - y^2)}{y^2} , \nnr
&& H = 1 + \frac{q}{r^2 + y^2} , \quad q = 2 m s^2 , \quad s = \sinh
\delta , \quad \mathcal{A} = \ud t' + y^2 \, \ud \psi_1 .
\eea
This particular example falls under the framework of the equal charge ungauged supergravity solutions dealt with in Section 3, since the additional parameter, the gauge-coupling constant $g$, enters the metric only via the functions $R$ and $Y$.

The Jordan frame metric, which has vielbeins $\tilde{e}^A = H^{-1/3} e^A$, has an irreducible KS tensor
\ben
\widetilde{K} = y^2 (- \tilde{e}^0 \tilde{e}^0 + \tilde{e}^1 \tilde{e}^1)
- r^2 (\tilde{e}^2 \tilde{e}^2 + \tilde{e}^3 \tilde{e}^3) + (y^2 -
r^2) \tilde{e}^4 \tilde{e}^4 .
\een
Returning to the Einstein frame metric, we therefore have an irreducible CKS tensor
\ben
Q = H^{2/3} [ y^2 (- e^0 e^0 + e^1 e^1) - r^2 (e^2 e^2 + e^3 e^3) + (y^2 -
r^2) e^4 e^4 ] ,
\een
which is not of gradient type.


\paragraph{Three equal charges, $\delta_1 = \delta_2 = \delta_3$:}


Another black hole solution of 5-dimensional $\uU (1)^3$ gauged supergravity has all 3 of its charge parameters equal, $\delta_1 = \delta_2 = \delta_3$ \cite{bh5dgsugra}.  It can be regarded as a solution of 5-dimensional minimal gauged supergravity.  In this case, the irreducible KS tensor of the uncharged solution, 5-dimensional Kerr--AdS, generalizes to the charged solution.  The Lorentzian metric has
\bea
e^0 & = & \sqrt{\frac{R}{r^2 + y^2}} \mathcal{A} , \quad e^1 =
\sqrt{\frac{r^2 + y^2}{R}} \, \ud r , \quad e^2 =
\sqrt{\frac{Y}{r^2 + y^2}} (\ud t' - r^2 \, \ud \psi_1) ,
\nnr
e^3 & = & \sqrt{\frac{r^2 + y^2}{Y}} \, \ud y , \quad e^4 =
\frac{a b}{r y} \left( \ud t' + (y^2 - r^2) \, \ud \psi_1 -
  r^2 y^2 \, \ud \psi_2 + \frac{q y^2}{a b (r^2 + y^2)} \mathcal{A}
\right) ,
\eea
and
\bea
&& R = \frac{(1 + g^2 r^2) (r^2 + a^2) (r^2 + b^2) + 2abq + q^2}{r^2} - 2m , \quad Y = - \frac{(1 - g^2 y^2) (a^2 - y^2) (b^2 - y^2)}{y^2} , \nnr
&& q = 2 m s^2 , \quad s = \sinh \delta , \quad \mathcal{A} = \ud t' + y^2 \, \ud \psi_1 .
\eea
We have an irreducible KS tensor
\ben
K = y^2 (- e^0 e^0 + e^1 e^1) - r^2 (e^2 e^2 + e^3 e^3) + (y^2 - r^2) e^4 e^4 ,
\een
which agrees with that found in \cite{symKerrAdS5}, up to a reducible KS tensor.


\paragraph{Two equal charges, third charge arbitrary, $\delta_1 = \delta_2$:}


There is a more general black hole solution \cite{newrotbh5d} of 5-dimensional $\uU (1)^3$ gauged supergravity that interpolates between these 2 solutions just discussed \cite{5dgsugrabhind, bh5dgsugra}, having two of the three charges equal, $\delta_1 = \delta_2$, and the third charge arbitrary.  The Lorentzian metric has
\bea
e^0 & = & H_1^{-2/3} H_3^{1/6} \sqrt{\frac{R}{r^2 + y^2}}
\frac{r}{\sqrt{r^2 + \gamma_r}} (\ud t' + y^2 \, \ud \psi_1) , \quad e^1 = H_1^{1/3} H_3^{1/6} \sqrt{\frac{r^2 + y^2}{R}} \, \ud r , \nnr
e^2 & = & H_1^{-2/3} H_3^{1/6} \sqrt{\frac{Y}{r^2 + y^2}}
\frac{y}{\sqrt{y^2 + \gamma_y}} [\ud t' - (r^2 + 2 m s_1^2) \, \ud
\psi_1] , \quad e^3 = H_1^{1/3} H_3^{1/6} \sqrt{\frac{r^2 + y^2}{Y}} \, \ud y , \nnr
e^4 & = & \frac{H_1^{1/3} H_3^{-1/3} a b}{\sqrt{(r^2 + \gamma_r) (y^2
    + \gamma_y)}} \bigg\{ \left[ r^2 + \gamma_r + \left( 1 + \frac{2 m
      s_3 c_3}{ab} \right) (y^2 + \gamma_y) \right] \frac{(\ud t' +
  y^2 \, \ud \psi_1)}{H_1 (r^2 + y^2)} \nnr
&& \qquad \qquad \qquad \qquad \qquad - (r^2 +
\gamma_r) [\ud \psi_1 + (y^2 + \gamma_y) \, \ud \psi_2] \bigg\} ,
\eea
and
\bea
&& R = \frac{(r^2 + \tilde{a}^2) (r^2 + \tilde{b}^2) + g^2 (r^2 +
  \tilde{a}^2 + 2 m s_1^2) (r^2 + \tilde{b}^2 + 2 m s_1^2) (r^2 +
  \gamma_r)}{r^2} - 2m , \nnr
&& Y = - \frac{[1 - g^2 (y^2 + \gamma_y)] (\tilde{a}^2 - y^2)
  (\tilde{b}^2 - y^2)}{y^2} , \nnr
&& \tilde{a} = a c_3 + b s_3 , \quad \tilde{b} = b c_3 + a s_3 , \quad \gamma_r = - \gamma_y + 2 m s_3^2 , \quad
\gamma_y = - [(a^2 + b^2) s_3^2 + 2 a b s_3 c_3] , \nnr
&& H_I = 1 + \frac{2 m s_I^2}{r^2 + y^2} , \quad s_I = \sinh \delta_I
, \quad c_I = \cosh \delta_I .
\eea
Compared with \cite{newrotbh5d}, we have made the notational changes $x \rightarrow r$, $y \rightarrow \ui y$, $t \rightarrow t'$, $\sigma \rightarrow - \psi_1$, $\chi \rightarrow - a b \psi_2$.

The Jordan frame metric, which has vielbeins $\tilde{e}^A = H_1^{-1/3} H_3^{-1/6} e^A$, has an irreducible KS tensor
\ben
\widetilde{K} = y^2 (-\tilde{e}^{0} \tilde{e}^0 + \tilde{e}^1 \tilde{e}^1)
- r^2 (\tilde{e}^2 \tilde{e}^2 + \tilde{e}^3 \tilde{e}^3) + \left(
  \frac{y^2 + \gamma_y}{H_3} - r^2 \right) \tilde{e}^4 \tilde{e}^4 .
\een
Returning to the Einstein frame metric, we therefore have an irreducible CKS tensor
\ben
Q = H_1^{2/3} H_3^{1/3} \left[ y^2 (-e^{0} e^0 + e^1 e^1) - r^2 (e^2 e^2 +
e^3 e^3) + \left( \frac{y^2 + \gamma_y}{H_3} - r^2 \right) e^4 e^4 \right] .
\een
Despite the complexity of the solution, it is noteworthy that we have such simple expressions for (C)KS tensors.

The inverse of the Jordan frame metric is
\bea
\left( \frac{\pd}{\pd \tilde{s}} \right) ^2 & = & - \frac{(1 + \gamma_r /
  r^2)}{(r^2 + y^2) R} \left[ (r^2 + 2 m s_1^2) \frac{\pd}{\pd t'} +
  \frac{\pd}{\pd \psi_1} + \left( 1 + \frac{2 m s_3 c_3}{a b} \right)
  \frac{1}{r^2 + \gamma_r} \frac{\pd}{\pd \psi_2} \right] ^2 \nnr
&& + \frac{R}{r^2 + y^2} \left( \frac{\pd}{\pd r} \right) ^2 +
\frac{(1 + \gamma_y / y^2)}{(r^2 + y^2) Y} \left( y^2 \frac{\pd}{\pd
    t'} - \frac{\pd}{\pd \psi_1} + \frac{1}{y^2 + \gamma_y}
  \frac{\pd}{\pd \psi_2} \right) ^2 \nnr
&& + \frac{Y}{r^2 + y^2} \left( \frac{\pd}{\pd y} \right) ^2 +
\frac{H_3}{a^2 b^2 (r^2 + \gamma_r) (y^2 + \gamma_y)} \left(
  \frac{\pd}{\pd \psi_2} \right) ^2 .
\eea
Since the components of $(r^2 + y^2) \tilde{g}^{ab}$ are additively separable as the sum of a function of $r$ and a function of $y$, we see directly that the HJE for geodesic motion on the Jordan frame metric separate.  For the Einstein frame metric, again the null HJE and massless KGE separate.


\paragraph{Further vacuum generalization:}


Returning to vacuum solutions, there is a more recent generalization \cite{newbh5d} of the 5-dimensional Myers--Perry metric, including the black ring with a single rotation parameter \cite{blackring} as a limiting case.  However, no CKS tensor is known for this generalization.


\subsection{Six dimensions}


In $D = 6$, there is an $\uSU (2)$ gauged supergravity theory, and a black hole solution with a single $\uU (1)$ charge was found in \cite{crotbh6}.  The Lorentzian metric has
\bea
e^0 & = & H^{-3/4} \sqrt{\frac{R}{(r^2 + y^2) (r^2 + z^2)}} \mathcal{A} , \quad e^1 = H^{1/4} \sqrt{\frac{(r^2 + y^2) (r^2 + z^2)}{R}} \, \ud r , \nnr
e^2 & = & H^{1/4} \sqrt{\frac{Y}{(r^2 + y^2) (y^2 - z^2)}} \left[ \ud t' + (z^2 - r^2) \, \ud \psi_1 - r^2 z^2 \, \ud \psi_2 - \frac{qr}{H (r^2 + y^2) (r^2 + z^2)} \mathcal{A} \right] , \nnr
e^3 & = & H^{1/4} \sqrt{\frac{(r^2 + y^2) (y^2 - z^2)}{Y}} \, \ud y , \nnr
e^4 & = & H^{1/4} \sqrt{\frac{Z}{(r^2 + z^2) (z^2 - y^2)}} \left[ \ud t' + (y^2 - r^2) \, \ud \psi_1 - r^2 y^2 \, \ud \psi_2 - \frac{qr}{H (r^2 + y^2) (r^2 + z^2)} \mathcal{A} \right] , \nnr
e^5 & = & H^{1/4} \sqrt{\frac{(r^2 + z^2) (z^2 - y^2)}{Z}} \, \ud z ,
\eea
and
\bea
&& R = (r^2 + a^2) (r^2 + b^2) + g^2 [r (r^2 + a^2) + q][r (r^2 + b^2) + q] - 2 m r , \nnr
&& Y = - (1 - g^2 y^2) (a^2 - y^2) (b^2 - y^2) , \quad Z = - (1 - g^2 z^2) (a^2 - z^2) (b^2 - z^2) , \nnr
&& H = 1 + \frac{qr}{(r^2 + y^2) (r^2 + z^2)} , \quad q = 2 m s^2 ,
\quad s = \sinh \delta , \nnr
&& \mathcal{A} = \ud t' + (y^2 + z^2) \, \ud \psi_1 + y^2 z^2 \, \ud \psi_2 .
\eea
The Jordan frame metric, which has vielbeins $\tilde{e}^A = H^{-1/4} e^A$, has irreducible KS tensors
\bea
\widetilde{K}^{(1)} & = & (y^2 + z^2) (- \tilde{e}^0 \tilde{e}^0 + \tilde{e}^1 \tilde{e}^1) + (z^2 - r^2) (\tilde{e}^2 \tilde{e}^2 + \tilde{e}^3 \tilde{e}^3) + (y^2 - r^2) (\tilde{e}^4 \tilde{e}^4 + \tilde{e}^5 \tilde{e}^5) , \nnr
\widetilde{K}^{(2)} & = & y^2 z^2 (- \tilde{e}^0 \tilde{e}^0 + \tilde{e}^1 \tilde{e}^1) - r^2 z^2 (\tilde{e}^2 \tilde{e}^2 + \tilde{e}^3 \tilde{e}^3) - r^2 y^2 (\tilde{e}^4 \tilde{e}^4 + \tilde{e}^5 \tilde{e}^5) .
\eea
Returning to the Einstein frame metric, we therefore have irreducible CKS tensors
\bea
Q^{(1)} & = & H^{1/2} [ (y^2 + z^2) (- e^0 e^0 + e^1 e^1) + (z^2 - r^2) (e^2 e^2 + e^3 e^3) + (y^2 - r^2) (e^4 e^4 + e^5 e^5) ] , \nnr
Q^{(2)} & = & H^{1/2} [ y^2 z^2 (- e^0 e^0 + e^1 e^1) - r^2 z^2 (e^2 e^2 + e^3 e^3) - r^2 y^2 (e^4 e^4 + e^5 e^5) ] ,
\eea
which are not of gradient type.


\subsection{Seven dimensions}


In $D = 7$, there is a $\uU (1)^2$ gauged supergravity theory, and a black hole solution with equal $\uU (1)$ charges was found in \cite{equalcharge}.  The CKS tensors of the ungauged solution generalize to the gauged solution.  The Lorentzian metric has
\bea
e^0 & = & H^{-4/5} \sqrt{\frac{R}{(r^2 + y^2) (r^2 + z^2)}} \mathcal{A} , \quad e^1 = H^{1/5} \sqrt{\frac{(r^2 + y^2) (r^2 + z^2)}{R}} \, \ud r , \nnr
e^2 & = & H^{1/5} \sqrt{\frac{Y}{(r^2 + y^2) (y^2 - z^2)}} \left[
  \ud t' + (z^2 - r^2) \, \ud \psi_1 - r^2 z^2 \, \ud
  \psi_2 - \frac{q}{H (r^2 + y^2) (r^2 + z^2)} \mathcal{A} \right] ,
\nnr
e^3 & = & H^{1/5} \sqrt{\frac{(r^2 + y^2) (y^2 - z^2)}{Y}} \, \ud y , \nnr
e^4 & = & H^{1/5} \sqrt{\frac{Z}{(r^2 + z^2) (z^2 - y^2)}} \left[
  \ud t' + (y^2 - r^2) \, \ud \psi_1 - r^2 y^2 \, \ud
  \psi_2 - \frac{q}{H (r^2 + y^2) (r^2 + z^2)} \mathcal{A} \right] ,
\nnr
e^5 & = & H^{1/5} \sqrt{\frac{(r^2 + z^2) (z^2 - y^2)}{Z}} \, \ud z, \nnr
e^6 & = & H^{1/5} \frac{a_1 a_2 a_3}{r y z} \bigg[ \ud t' + (y^2 + z^2
- r^2) \, \ud \psi_1 + (y^2 z^2 - r^2 y^2 - r^2 z^2) \, \ud
\psi_2 - r^2 y^2 z^2 \, \ud \psi_3 \nnr
&& \qquad \qquad \qquad - \frac{q}{H (r^2 + y^2) (r^2 + z^2)} \bigg( 1 +
\frac{g y^2 z^2}{a_1 a_2 a_3} \bigg) \mathcal{A} \bigg] ,
\eea
and
\bea
&& R = \frac{1 + g^2 r^2}{r^2} \prod_{i=1}^3 (r^2 + a_i^2) + q
g^2 ( 2 r^2 + a_1^2 + a_2^2 + a_3^2 ) - \frac{2 q g a_1 a_2 a_3}{r^2} +
\frac{q^2 g^2}{r^2} - 2m , \nnr
&& Y = \frac{1 - g^2 y^2}{y^2} \prod_{i=1}^3 (a_i^2 - y^2) , \quad Z =
\frac{1 - g^2 z^2}{z^2} \prod_{i=1}^3 (a_i^2 - z^2) , \quad H = 1 + \frac{q}{(r^2 + y^2) (r^2 + z^2)} , \nnr
&& q = 2 m s^2 , \quad s = \sinh \delta
, \quad \mathcal{A} = \ud t' + (y^2 + z^2) \, \ud \psi_1 + y^2 z^2 \, \ud \psi_2 .
\eea
The Jordan frame metric, which has vielbeins $\tilde{e}^A = H^{-1/5} e^A$, has irreducible KS tensors
\bea
\widetilde{K}^{(1)} & = & (y^2 + z^2) (- \tilde{e}^0 \tilde{e}^0 +
\tilde{e}^1 \tilde{e}^1) + (z^2 - r^2) (\tilde{e}^2 \tilde{e}^2 +
\tilde{e}^3 \tilde{e}^3) + (y^2 - r^2) (\tilde{e}^4 \tilde{e}^4 +
\tilde{e}^5 \tilde{e}^5) \nnr
&& + (y^2 + z^2 - r^2) \tilde{e}^6 \tilde{e}^6 , \nnr
\widetilde{K}^{(2)} & = & y^2 z^2 (- \tilde{e}^0 \tilde{e}^0 + \tilde{e}^1
\tilde{e}^1) - r^2 z^2 (\tilde{e}^2 \tilde{e}^2 + \tilde{e}^3
\tilde{e}^3) - r^2 y^2 (\tilde{e}^4 \tilde{e}^4 + \tilde{e}^5
\tilde{e}^5) \nnr
&& + (y^2 z^2 - r^2 y^2 - r^2 z^2) \tilde{e}^6 \tilde{e}^6 .
\eea
Returning to the Einstein frame metric, we therefore have irreducible CKS tensors
\bea
Q^{(1)} & = & H^{2/5} [(y^2 + z^2) (- e^0 e^0 + e^1 e^1) + (z^2 - r^2) (e^2 e^2 + e^3 e^3) + (y^2 - r^2) (e^4 e^4 + e^5 e^5) \nnr
&& \qquad + (y^2 + z^2 - r^2) e^6 e^6 ] , \nnr
Q^{(2)} & = & H^{2/5} [y^2 z^2 (- e^0 e^0 + e^1 e^1) - r^2 z^2 (e^2 e^2 + e^3 e^3) - r^2 y^2 (e^4 e^4 + e^5 e^5) \nnr
&& \qquad + (y^2 z^2 - r^2 y^2 - r^2 z^2) e^6 e^6 ] ,
\eea
which are not of gradient type.

We have seen that there is a separability structure for this solution, and it is straightforward to check directly the separation of the HJE for the Jordan frame metric, the null HJE and massless KGE for the Einstein frame metric.  The only adjustment to take into account, compared to Section 3.4 with $m_1 = m$ and $m_2 = m_3 = 0$, is an inverse vielbein of the analytically continued conformally related metric,
\ben
\tilde{e}{^{\hat{1}}} {^a} \pd_a = \frac{1}{\sqrt{X_1 U_1}} \left(
  \sum_{k=0}^3 (-x_1^2)^{2-k} \frac{\pd}{\pd \psi_k} + 2 m s^2
  \frac{\pd}{\pd \psi_0} + \frac{2 m s^2 g}{a_1 a_2 a_3 x_1^2}
  \frac{\pd}{\pd \psi_3} \right) ,
\een
and then, \emph{mutatis mutandis}, the separation occurs in the same manner.  


\section{Some unequal charge supergravity black holes}


Killing tensors for supergravity black holes without the equal charge simplification tend to be rather more complicated.  One exception is if all of the angular momenta are set equal, since the rotational symmetry group of the solution is enhanced, and there are additional Killing vectors that make the system completely integrable \cite{sepHJKGKdS}, so all KS tensors are reducible.

There are CKS tensors for some unequal charge supergravity black holes that have been studied in the literature.  Some of the features of the equal charge cases continue to hold.  In particular there is a conformally related Jordan frame metric with enough KS tensors, giving constants of motion in involution, to render geodesic motion completely integrable in the sense of Liouville for the Jordan frame metric.  Furthermore, there is a separability structure for the Jordan frame metric, so the HJE for the Jordan frame metric separates, and so the null HJE for the Einstein frame metric separates.


\subsection{Four dimensions}


Some initial understanding of the unequal charge situation comes by studying black holes in 4-dimensional ungauged supergravity.  The simplest such example is the 2-charge Cveti\v{c}--Youm solution, from which we can generalize to the 4-charge Cveti\v{c}--Youm solution.  With these examples, we find that we can simplify the solutions in a fairly algorithmic way.

An irreducible KS tensor can be read off from the separability of the HJE.  Such a tensor is not unique, since one can add a reducible KS tensor.  There is, however, a privileged KS tensor.  One can regard $K{_a}{^b}$, with mixed indices, as a linear map.  It has eigenvectors, which we shall refer to as eigenforms, i.e.~$\mathcal{E} = \mathcal{E}_a \ud x^a$ satisfying
\ben
K{_a}{^b} \mathcal{E}_b = \lambda \mathcal{E}_a , 
\een
for some eigenvalue $\lambda$.  In general the eigenvalues will all be distinct, however for the ungauged 4-dimensional black holes with independent $\uU (1)$ charges there is a unique (up to the addition of a constant multiple of the metric) irreducible KS tensor for which there is a repeated eigenvalue and which has 2 pairs of repeated eigenvalues in the uncharged limit.  This privileged KS tensor has 4 eigenforms, and it is natural to write the metric in terms of these privileged 1-forms.  The result in this 4-dimensional case is  that the metric has no cross-terms when written in terms of these eigenforms, i.e.~one can choose vielbeins to be eigenforms.

We shall start by presenting the 4-dimensional metric in terms of these eigenforms, since ultimately one wishes to take the metric as the starting point, but we should bear in mind that this form of the metric was derived by a systematic procedure.  To summarize, the procedure for obtaining the form of the metric is as follows:
\begin{enumerate}
\item Consider the Jordan frame metric $\ud \tilde{s}^2$, for simplicity.
\item Read off an irreducible KS tensor $\widetilde{K}^{ab}$ from the separability of the HJE.
\item Add a constant multiple of the inverse metric $\tilde{g}^{ab}$ to modify $\widetilde{K}^{ab}$ so that its part involving $\ud r^2$ and $\ud y^2$ is symmetric under the interchange of $x^2 = - r^2$ and $y^2$, including interchanging $X(x) = R(r)$ and $Y(y)$; this is convenient as it reproduces the usual symmetric KS tensor of the uncharged solution.
\item Add symmetrized products of the Killing vectors so that $-y^2$ is a repeated eigenvalue, and with the correct uncharged limit.
\item Compute the eigenforms for these eigenvalues.
\item Write the metric in terms of these eigenforms.
\item Transfer the results to the Einstein frame metric $\ud s^2$.
\end{enumerate}


\subsubsection{2-charge Cveti\v{c}--Youm solution}


The simplest example to illustrate some features that are different for solutions with unequal charges is the 2-charge Cveti\v{c}--Youm solution in 4 dimensions with both $\uU (1)$ charges arbitrary.  The Lorentzian metric has
\bea
e^0 & = & (H_1 H_2)^{-1/4} \left( 1 + \frac{2 m r (c_1 - c_2)^2 (a^2 - y^2)}{H_1 H_2 (r^2 + y^2)^2} \right) ^{-1/2} \sqrt{\frac{R}{r^2 + y^2}} \left( \ud t - \frac{a^2 - y^2}{a} \, \ud \phi \right) , \nnr
e^1 & = & (H_1 H_2)^{1/4} \sqrt{\frac{r^2 + y^2}{R}} \, \ud r , \nnr
e^2 & = & (H_1 H_2)^{1/4} \left( 1 + \frac{2 m r (c_1 - c_2)^2 (a^2 - y^2)}{H_1 H_2 (r^2 + y^2)^2} \right) ^{1/2} \sqrt{\frac{Y}{r^2 + y^2}} \bigg[ \ud t - \frac{r^2 + a^2}{a} \, \ud \phi \nnr
&& - \left( 1 - \frac{1 + 2 m r (c_1 c_2 - 1) / (r^2 + y^2)}{H_1 H_2 [ 1 + 2 m r (c_1 - c_2)^2 (a^2 - y^2) / H_1 H_2 (r^2 + y^2)^2 ]} \right) \left( \ud t - \frac{a^2 - y^2}{a} \, \ud \phi \right) \bigg] , \nnr
e^3 & = & (H_1 H_2)^{1/4} \sqrt{\frac{r^2 + y^2}{Y}} \, \ud y ,
\eea
and
\ben
R = r^2 + a^2 - 2 m r , \quad Y = a^2 - y^2 , \quad H_I = 1 + \frac{2 m s_I^2 r}{r^2 + y^2} , \quad s_I = \sinh \delta_I , \quad c_I = \cosh \delta_I .
\een
The Jordan frame metric, which has vielbeins $\tilde{e}^A = (H_1 H_2)^{-1/4} e^A$, has an irreducible KS tensor
\ben
\widetilde{K} = y^2 (- \tilde{e}^0 \tilde{e}^0 + \tilde{e}^1 \tilde{e}^1) - \left( r^2 + \frac{2 m r (c_1 - c_2)^2 (a^2 - y^2)}{H_1 H_2 (r^2 + y^2)} \right) \tilde{e}^2 \tilde{e}^2 - r^2 \tilde{e}^3 \tilde{e}^3 .
\een
We see that the vielbeins $\tilde{e}^A$ are eigenforms of $\widetilde{K}$.  Returning to the Einstein frame metric, we therefore have an irreducible CKS tensor
\ben
Q = (H_1 H_2)^{1/2} \left[ y^2 (- e^0 e^0 + e^1 e^1) - \left( r^2 + \frac{2 m r (c_1 - c_2)^2 (a^2 - y^2)}{H_1 H_2 (r^2 + y^2)} \right) e^2 e^2 - r^2 e^3 e^3 \right] .
\een


\subsubsection{4-charge Cveti\v{c}--Youm solution}


The 4-charge Cveti\v{c}--Youm solution \cite{entropycrotbhst, crotbh4d} has all four $\uU (1)$ charges arbitrary.  The Lorentzian metric has
\bea
e^0 & = & \sqrt{\frac{W R}{W^2 + (a^2 - y^2) V}} \left( \ud t - \frac{a^2 - y^2}{a} \, \ud \phi \right) , \quad e^1 = \sqrt{\frac{W}{R}} \, \ud r , \nnr
e^2 & = & \sqrt{\frac{W^2 + (a^2 - y^2) V}{W (r^2 + y^2)}} \sqrt{\frac{Y}{r^2 + y^2}} \bigg[ \ud t - \frac{r^2 + a^2}{a} \, \ud \phi \nnr
&& - \left( 1 - \frac{(r^2 + y^2) [ r^2 + y^2 + 2 m r (c_{1234} - s_{1234} - 1) + 4 m^2 s_{1234}^2 ] }{W^2 + (a^2 - y^2) V} \right) \left( \ud t - \frac{a^2 - y^2}{a} \, \ud \phi \right) \bigg] , \nnr
e^3 & = & \sqrt{\frac{W}{Y}} \, \ud y ,
\eea
and
\bea
&& R = r^2 + a^2 - 2 m r , \quad Y = a^2 - y^2 , \nnr
&& V = 2 m r [(c_{13} - c_{24})^2 - (s_{13} -
s_{24})^2] + 4 m^2 [(s_{13} - s_{24})^2 -
(s_{13} c_{24} - s_{24} c_{13})^2 ] , \nnr
&& W^2 = (r_1 r_3 + y^2) (r_2 r_4 + y^2) - 4 m^2 (s_{13} c_{24} - s_{24} c_{13})^2 y^2 , \quad r_I = r + 2 m s_I^2 , \nnr
&& s_{I \ldots J} = s_I \ldots s_J , \quad c_{I \ldots J} = c_I \ldots c_J , \quad s_I = \sinh \delta_I , \quad c_I = \cosh \delta_I .
\eea
We have presented $V$ and $W^2$ here in a concise manner that makes clearer what simplification occurs when certain charges are set equal, but they can be presented so that they are manifestly symmetric in the indices $I = 1, 2, 3, 4$.

The Jordan frame metric, which has vielbeins $\te^A = \sqrt{(r^2 + y^2) / W} e^A$, has an irreducible KS tensor
\ben
\widetilde{K} = y^2 ( - \te^0 \te^0 + \te^1 \te^1) - \left( r^2 + \frac{(r^2 + y^2) (a^2 - y^2) V}{W^2} \right) \te^2 \te^2 - r^2 \te^3 \te^3 .
\een
We again see that $\mathcal{E}_1$ and $\mathcal{E}_2$ are eigenforms of $\widetilde{K}$.  Returning to the Einstein frame metric, we therefore have an irreducible CKS tensor
\ben
Q = \frac{W}{r^2 + y^2} \left[ y^2 ( - e^0 e^0 + e^1 e^1) - \left( r^2 + \frac{(r^2 + y^2) (a^2 - y^2) V}{W^2} \right) e^2 e^2 - r^2 e^3 e^3 \right] .
\een

$-r^2$ is clearly a repeated eigenvalue of $\widetilde{K}$ if the charges are pairwise equal.  We can also prove the converse: if $-r^2$ is a repeated eigenvalue for this privileged KS tensor, then the charges are pairwise equal.  From the expression for $V$, there are 2 constraints on the charge parameters $\delta_I$, coming from the $m$ and $m^2$ coefficients.  We therefore expect that solutions with the desired property are specified by 2, rather than 4, charge parameters.  Using elementary identities for hyperbolic functions, and helped by writing the $m$ coefficient of $V$ as a difference of two squares, the constraint that arises from the $m$ coefficient of $V$ may be expressed as
\ben
[ \cosh (\delta_1 - \delta_3) - \cosh (\delta_2 - \delta_4) ] [ \cosh (\delta_1 + \delta_3) - \cosh (\delta_2 + \delta_4) ] = 0 .
\een
Again helped by a difference of two squares, the constraint that arises from the $m^2$ coefficient of $V$ may be expressed as
\bea
&& \{ [1 + \cosh(\delta_1 + \delta_3)] [1 + \cosh (\delta_2 - \delta_4)] - [1 + \cosh (\delta_2 + \delta_4)] [1 + \cosh(\delta_1 - \delta_3)] \} \nnr
&& \times \{ [1 - \cosh (\delta_1 - \delta_3)] [1 - \cosh (\delta_2 + \delta_4)] - [1 - \cosh (\delta_2 - \delta_4)] [1 - \cosh (\delta_1 + \delta_3)] \} = 0 . \nnr
\eea
It is now easy to see that a repeated $-r^2$ eigenvalue is equivalent to pairwise equal charges.  From the general 4-charge metric, the simplification to pairwise equal charges has emerged in a natural way from purely geometric considerations, by considering the eigenvalues of a privileged KS tensor.


\subsection{Five dimensions}


We have not yet been able to present the general 3-charge Cveti\v{c}--Youm solution in 5 dimensions in a similar manner to the 4-charge Cveti\v{c}--Youm solution in 4 dimensions.  Here, we collect together some further results for examples from 5 dimensions.


\subsubsection{3-charge Cveti\v{c}--Youm solution}


The general 3-charge Cveti\v{c}--Youm solution \cite{rot5dbhhet} of $D = 5$ ungauged supergravity does not fall within the equal charge simplification that we focus on in this paper.  However, an irreducible CKS tensor is known \cite{sepmcbh}, which again is induced by a KS tensor for a conformally related Jordan frame metric.  To be more consistent with the rest of this paper, we alter notation slightly: compared with \cite{sepmcbh}, we define $y^2 = a^2 \cos^2 \theta + b^2 \sin^2 \theta$, $s_1 = \textrm{sh}_{e1}$, $s_2 = \textrm{sh}_{e2}$, $s_3 = \textrm{sh}_e$, $H_I = 1 + 2 m s_I^2 / (r^2 + y^2)$, and so $\bar{\Delta} = H_1 H_2 H_3 (r^2 + y^2)^3$.

Denoting the Einstein frame metric by $\ud s^2$, we consider the Jordan frame metric $\ud \tilde{s}^2 = (H_1 H_2 H_3)^{-1/3} \, \ud s^2$.  An irreducible KS tensor for the Jordan frame metric is $\widetilde{K}^{ab} = \widetilde{Q}^{ab} - y^2 \tilde{g}^{ab}$, where
\ben
\widetilde{Q}^{ab} \, \pd_a \, \pd_b = - y^2 \left( \frac{\pd}{\pd t} \right) ^2 + \frac{a^2 - b^2}{a^2 -
  y^2} \left( \frac{\pd}{\pd \phi} \right) ^2 + \frac{b^2 - a^2}{b^2 - y^2}
\left( \frac{\pd}{\pd \psi} \right) ^2 - \frac{(a^2 - y^2) (b^2 - y^2)}{y^2}
\left( \frac{\pd}{\pd y} \right) ^2 .
\een
Returning to the Einstein frame metric, we have an induced irreducible CKS tensor with its contravariant components given by $Q^{ab} = \widetilde{Q}^{ab}$, which is the CKS tensor presented in \cite{sepmcbh}.


\subsubsection{Supersymmetric gauged supergravity black holes}


Supersymmetric black holes of $D = 5$, $\uU (1)^3$ gauged supergravity with 2 arbitrary angular momenta and 3 arbitrary $\uU (1)$ charges, except for one constraint, were obtained in \cite{susymcAdS5}.  The solution in general has unequal $\uU (1)$ charges, unlike the gauged solutions considered above, however one can choose a simple set of vielbeins, and so there are some similarities to what happens in the equal charge cases.  We shall again find that there is a CKS tensor for\ the Einstein frame metric that is induced by a KS tensor for the Jordan frame metric, and that these tensors have fairly simple expressions in terms of a certain choice of vielbeins.

The background theory behind supersymmetric black hole solutions of 5-dimensional $\uU (1)^3$ gauged supergravity is in \cite{5dgaugedsugra, gensusyAdS5bh}.  For a supersymmetric solution, there is a Killing spinor $\epsilon$, so one can construct a Killing vector $K^a = \bar{\epsilon} \gamma^a \epsilon$ that is timelike or null.  The black hole solutions belong to the timelike case, for which the metric takes a canonical form,
\ben
\ud s^2 = - f^2 (\ud t + \omega)^2 + f^{-1} \, \ud \bar{s}_4^2 ,
\een
where $\ud \bar{s}_4^2$ is a 4-dimensional K\"{a}hler metric.  One can choose vielbeins for the K\"{a}hler base metric so that its K\"{a}hler form is $J = \bar{e}^1 \wedge \bar{e}^2 + \bar{e}^3 \wedge \bar{e}^4$.  For this particular black hole solution, the K\"{a}hler base metric is \cite{susymcAdS5}
\ben
\ud \bar{s}_4^2 = \Delta_r \left( \sin^2 \theta \frac{\ud \phi}{\Xi_a} + \cos^2 \theta \frac{\ud \psi}{\Xi_b} \right) ^2 + \frac{r^2}{\Delta_r} \, \ud r^2 + r^2 \sin^2 \theta \cos^2 \theta \, \Delta_\theta \left( \frac{\ud \phi}{\Xi_a} - \frac{\ud \psi}{\Xi_b} \right) ^2 + \frac{r^2}{\Delta_\theta}\,  \ud \theta^2 ,
\een
where
\ben
\Delta_r = g^2 r^4 + (1 + ag + bg)^2 r^2 , \quad \Delta_\theta = 1 - a^2 g^2 \cos^2 \theta - b^2 g^2 \sin^2 \theta , \quad \Xi_a = 1 - a^2 g^2 , \quad \Xi_b = 1 - b^2 g^2 .
\een
It belongs to the orthotoric family of K\"{a}hler metrics \cite{geomwSDKah, Ham2formsI}.  Such orthotoric metrics can be obtained from a limit of the higher-dimensional Kerr-type metric \cite{KNdScurv}; in 4 (real) dimensions, this limiting procedure was originally done by starting with the euclidean Pleba\'{n}ski--Demia\'{n}ski metric \cite{toricSES2S3}.  Coming from generalizations of the Kerr metric, one can choose vielbeins with a similar structure as those for the Kerr metric, as can be read off from the way we have presented it above.  Its K\"{a}hler form is
\ben
J = \frac{1}{2} \ud \left[ r^2 \left( \sin^2 \theta \frac{\ud \phi}{\Xi_a} + \cos^2 \theta \frac{\ud \psi}{\Xi_b} \right) \right] .
\een

The 5-dimensional Lorentzian metric has
\bea
&& e^0 = (H_1 H_2 H_3)^{-1/3} (\ud t + \omega_\phi \, \ud \phi + \omega_\psi \, \ud \psi) , \nnr
&& e^1 = (H_1 H_2 H_3)^{1/6} \frac{r}{\sqrt{R}} \, \ud r , \quad e^2 =
(H_1 H_2 H_3)^{1/6} \sqrt{R} \left( \frac{(a^2 - y^2) \, \ud \phi}{\Xi_a
    (a^2 - b^2)} + \frac{(b^2 - y^2) \, \ud \psi}{\Xi_b (b^2 - a^2)}
\right) , \nnr
&& e^3 = (H_1 H_2 H_3)^{1/6} \frac{r}{\sqrt{Y}} \, \ud y , \quad e^4 =
(H_1 H_2 H_3)^{1/6} r y \sqrt{Y} \left( \frac{\ud \phi}{\Xi_a (a^2 -
    b^2)} + \frac{\ud \psi}{\Xi_b (b^2 - a^2)} \right) , \nnr
\eea
with
\bea
\omega_\phi & = & - \frac{g (a^2 - y^2)}{\Xi_a (a^2 - b^2) r^2} \bigg[
(r^2 + y^2)^2 + \left( a^2 + 2ab + \frac{2 (a+b)}{g}\right) (r^2 +
y^2) \nnr
&& \qquad \qquad \qquad \quad + \frac{\nu_1 \nu_2 + \nu_2 \nu_3 + \nu_3 \nu_1}{2 g^4} - \frac{a^2 b^2}{2} + \frac{b^2 - a^2}{2 g^2} \bigg] , \nnr
\omega_\psi & = & - \frac{g (b^2 - y^2)}{\Xi_b (b^2 - a^2) r^2} \bigg[
(r^2 + y^2)^2 + \left( b^2 + 2ab + \frac{2 (a+b)}{g}\right) (r^2 +
y^2) \nnr
&& \qquad \qquad \qquad \quad + \frac{\nu_1 \nu_2 + \nu_2 \nu_3 + \nu_3 \nu_1}{2 g^4} - \frac{a^2 b^2}{2} + \frac{a^2 - b^2}{2 g^2} \bigg] ,
\eea
and
\ben
R = g^2 r^4 + (1 + ag + bg)^2 r^2 , \quad Y = - \frac{(1 - g^2 y^2) (a^2 - y^2) (b^2 - y^2)}{y^2} , \quad H_I = 1 + \frac{g^2 y^2 + \nu_I}{g^2 r^2} .
\een
Note that the coordinates used here differ from those used in the other examples of this paper; most significantly, the timelike Killing vector, which is obtained from a Killing spinor, does not coincide with the time coordinates of the other examples.  Compared with \cite{susymcAdS5}, we have defined $\nu_I = \sqrt{\Xi_a \Xi_b} (1 + g^2 \mu_I) - 1$, which simplifies the expressions slightly, and, as elsewhere in this paper, we have $y^2 = a^2 \cos^2 \theta + b^2 \sin^2 \theta$.  There is a constraint on the parameters $\nu_I$:
\ben
\nu_1 + \nu_2 + \nu_3 = 2 ( ag + bg + a b g^2 ) .
\een

Considering the Jordan frame metric for which we have vielbeins $\tilde{e}^A = (H_1 H_2 H_3)^{-1/6} e^A$, there is an irreducible KS tensor,
\bea
\widetilde{K} & = & \frac{(1 - g^2 y^2) [1 - (a^2 - y^2) (b^2 - y^2) / r^4]}{H_1 H_2 H_3 g^2} \tilde{e}^0 \tilde{e}^0 \nnr
&& + \frac{1}{(H_1 H_2 H_3)^{1/2} g r} [r \sqrt{R} (\tilde{e}^0
\tilde{e}^2 + \tilde{e}^2 \tilde{e}^0) + y \sqrt{Y} (\tilde{e}^0
\tilde{e}^4 + \tilde{e}^4 \tilde{e}^0) ] + r^2 (\tilde{e}^3
\tilde{e}^3 + \tilde{e}^4 \tilde{e}^4) .
\eea
Returning to the Einstein frame metric, we therefore have an irreducible CKS tensor
\bea
Q & = & \frac{(1 - g^2 y^2) [1 - (a^2 - y^2) (b^2 - y^2) / r^4]}{H_1 H_2 H_3 g^2} e^0 e^0 \nnr
&& + \frac{1}{(H_1 H_2 H_3)^{1/2} g r} [r \sqrt{R} (e^0 e^2 + e^2 e^0) + y \sqrt{Y} (e^0 e^4 + e^4 e^0) ] + r^2 (e^3 e^3 + e^4 e^4) ,
\eea
which is in general not of gradient type.  The CKS tensor presented here agrees with that of \cite{sepmcbh} up to a reducible CKS tensor\footnote{One should note a typographical error in the expression for $r^2 H^{1/3} g^{t t}$ given in \cite{sepmcbh}, which should have the first term omitted, so the first of the 2 terms in the expression for $K^{tt}$ given there should be omitted.}.

However, the expressions in terms of vielbeins for the (C)KS tensors that we have given here are not as simple as in the other cases that we have studied.  There are cross terms, with non-zero coefficients for $e^0 e^2 + e^2 e^0$ and $e^0 e^4 + e^4 e^0$, and these cross terms cannot be removed by adding reducible (C)KS tensors.  This goes to emphasise the fact that the choice of vielbeins that we have used, which are adapted to the K\"{a}hler structure that arises from being a supersymmetric solution, does not coincide with the vielbeins that one usually uses for the higher-dimensional Kerr--NUT--AdS solution.


\section{Discussion}


We have considered the separability of the HJE and KGE for various metrics that arise from supergravity black holes, which we have seen is related to the existence of (C)KS tensors.  Such separation might furthermore be related to the existence of commuting symmetry operators, as has been found for the higher-dimensional Kerr--NUT--AdS metric \cite{symKGKNUTAdS}.

One might next consider the the Dirac equation, which separates for the higher-dimensional Kerr--NUT--AdS metric \cite{sepDiracKNUTdS}.  However, the separability of the Dirac equation appears to be related to the existence of (conformal) Killing--Yano tensors \cite{Diraceq}, and these do not seem to exist for the metrics arising from charged solutions that we have considered here, so separability might not be expected.

We have seen that many of the known supergravity black holes have metrics that are conformally related to metrics with sufficient (hidden) symmetry that geodesic motion on the conformally related metric is completely integrable in the sense of Liouville, and furthermore that this conformally related metric possesses a separability structure.  There are still black hole solutions of gauged supergravity theories with arbitrary angular momenta and charges that are yet to be discovered.  We expect that these more general metrics are also conformally related to metrics with separability structures, like all of the examples considered here.

For simplicity, we have not attempted any detailed analysis of the higher-dimensional charged Kerr--NUT solution for which the two $\uU (1)$ charges are unequal.  There is no known simplification of the solution in this more general case, and this is a substantial obstacle to generalizing to gauged supergravity.  We have, however, seen some simplification in the 4-dimensional case, and this might serve as a guide to such generalizations.


\section*{Acknowledgements}


I would like to thank Chris Pope for reading through a draft of this manuscript.  This work has been supported by STFC.


\appendix



\section{Higher-dimensional Kerr--NUT with two charges}


We present here the general 2-charge solution.  The analytically continued Jordan frame metric $\ud \tilde{s}^2$, related to the Einstein frame metric $\ud s^2$ by $\ud s^2 = (H_1 H_2)^{1/(D-2)} \, \ud \tilde{s}^2$, is
\bea
\ud \tilde{s}^2 & = & \sum_\mu \frac{U_\mu}{X_\mu} \, \ud x_\mu^2 - \bigg( 1 - \sum_\mu  \frac{2 N_\mu}{U_\mu} \bigg)
\frac{\ud t^2}{H_1 H_2} - c_1 c_2 \sum_i \sum_\mu \frac{2 N_\mu
  \tilde{\gamma}_i}{z_{i \mu}^2 U_\mu} \frac{2 \, \ud t \, \ud
  \tilde{\phi}_i}{H_1 H_2} \nnr
&& + \sum_i \bigg[ \frac{H_1 H_2 B_i}{\tilde{\gamma}_i} + \sum_\mu
\left( \frac{2 N_\mu}{U_\mu z_{i \mu}^4} - \frac{4 N_\mu^2 s_1^2
  s_2^2}{U_\mu^2 z_{i \mu}^4} \right) \nnr
&& \qquad \quad + \sum_{\mu < \nu} \left( \frac{4 N_\mu N_\nu (s_1^2 + s_2^2)
  x_{\mu \nu}^4}{U_\mu U_\nu z_{i \mu}^4 z_{i
    \nu}^4} - \frac{8 N_\mu N_\nu s_1^2
  s_2^2}{U_\mu U_\nu z_{i \mu}^2 z_{i \nu}^2} \right) + \sum_{\mu \neq \nu} \frac{8 N_\mu^2 N_\nu s_1^2
  s_2^2 x_{\mu \nu}^4}{U_\mu^2 U_\nu z_{i \mu}^4 z_{i \nu}^4} \nnr
&& \qquad \quad - \sum_{\mu < \nu < \rho} \frac{16 N_\mu N_\nu N_\rho
  s_1^2 s_2^2}{U_\mu U_\nu U_\rho z_{i \mu}^2 z_{i \nu}^2 z_{i \rho}^2} \left( \frac{x_{\nu \mu}^2 x_{\rho \mu}^2}{z_{i
    \mu}^2} + \frac{x_{\rho \nu}^2 x_{\mu \nu}^2}{z_{i \nu}^2} +
\frac{x_{\mu \rho}^2 x_{\nu \rho}^2}{z_{i \rho}^2} \right) \bigg]
\tilde{\gamma}_i^2 \frac{\ud \tilde{\phi}_i^2}{H_1 H_2} \nnr
&& + \sum_{i < j} \bigg[ \sum_\mu \left( \frac{2 N_\mu}{z_{i \mu}^2 z_{j \mu}^2 U_\mu} - \frac{4
  N_\mu^2 s_1^2 s_2^2}{z_{i \mu}^2
  z_{j \mu}^2 U_\mu^2} \right) + \sum_{\mu < \nu} \frac{4 N_\mu N_\nu (s_1^2 +
  s_2^2) x_{\mu \nu}^4}{z_{i \mu}^2
  z_{i \nu}^2 z_{j \mu}^2 z_{j \nu}^2 U_\mu U_\nu} \nnr
&& \qquad \quad - \sum_{\mu < \nu} \frac{4 N_\mu N_\nu s_1^2 s_2^2}{U_\mu U_\nu} \left( \frac{1}{z_{i
    \mu}^2 z_{j \nu}^2} + \frac{1}{z_{i \nu}^2 z_{j \mu}^2} \right) +
\sum_{\mu \neq \nu} \frac{8 N_\mu^2 N_\nu s_1^2 s_2^2 x_{\mu \nu}^4}{z_{i \mu}^2 z_{i \nu}^2 z_{j \mu}^2
  z_{j \nu}^2 U_\mu^2 U_\nu} \nnr
&& \qquad \quad + \sum_{\mu < \nu < \rho} \frac{8 N_\mu N_\nu N_\rho
  s_1^2 s_2^2}{z_{i \mu}^2 z_{i
    \nu}^2 z_{i \rho}^2 z_{j \mu}^2 z_{j \nu}^2 z_{j \rho}^2 U_\mu
  U_\nu U_\rho} [(z_{i \nu}^2 z_{j \rho}^2 + z_{i \rho}^2 z_{j \nu}^2)
x_{\nu \mu}^2 x_{\rho \mu}^2 \nnr
&& \qquad \qquad \qquad \qquad + (z_{i \rho}^2 z_{j \mu}^2 + z_{i
  \mu}^2 z_{j \rho}^2) x_{\rho \nu}^2 x_{\mu \nu}^2 + (z_{i \mu}^2
z_{j \nu}^2 + z_{i \nu}^2 z_{j \mu}^2) x_{\mu \rho}^2 x_{\nu \rho}^2]
\bigg] \tilde{\gamma}_i \tilde{\gamma_j} \frac{2 \, \ud \tilde{\phi}_i \, \ud \tilde{\phi}_j}{H_1 H_2} , \nnr
\label{2chargemetric}
\eea
and the other fields are
\bea
&& A_{(1)}^1 = \sum_\mu
\frac{2 N_\mu s_1}{H_1 U_\mu} \bigg( c_1 \, \ud t - c_2 \sum_i
\frac{\tilde{\gamma}_i}{z_{i \mu}^2} \, \ud
\tilde{\phi}_i \bigg) , \quad A_{(1)}^2 = \sum_\mu \frac{2 N_\mu s_2}{H_2 U_\mu} \bigg( c_2 \, \ud t - c_1 \sum_i
\frac{\tilde{\gamma}_i}{z_{i \mu}^2} \, \ud \tilde{\phi}_i \bigg) , \nnr
&& X_I = \frac{(H_1 H_2)^{(D-3)/2(D-2)}}{H_I}, \quad B_{(2)} = \sum_\mu \frac{N_\mu s_1 s_2}{U_\mu} \left(
\frac{1}{H_1} + \frac{1}{H_2} \right) \ud t \wedge \sum_i
\frac{\tilde{\gamma}_i}{z_{i \mu}^2} \, \ud \tilde{\phi}_i ,
\label{unequal1}
\eea
where for spacetime dimensions $D = 2 n + \varepsilon$, $\varepsilon = 0, 1$,
\bea
&& U_\mu = \sideset{}{'} \prod_{\nu = 1}^n (x_\nu^2 - x_\mu^2) , \quad X_\mu = - \frac{1}{(- x_\mu^2)^\varepsilon} \prod_{k=1}^{n-1+\varepsilon} (a_k^2 - x_\mu^2) + 2 N_\mu , \quad x_{\mu \nu}^2 = x_\mu^2 - x_\nu^2 , \nnr
&& H_I = 1 + \sum_{\mu = 1}^n \frac{2 N_\mu s_I^2}{U_\mu} , \quad N_\mu = m_\mu x_\mu^{1 - \varepsilon} , \quad s_I = \sinh \delta_I , \quad c_I = \cosh \delta_I , \nnr
&& \tilde{\gamma}_i = a_i^{2 \varepsilon} \prod_{\mu = 1}^n (a_i^2 - x_\mu^2) , \quad z_{i \mu}^2 = a_i^2 - x_\mu^2 , \quad B_i = \sideset{}{'} \prod_{k=1}^{n-1+\varepsilon} (a_i^2 - a_k^2) .
\eea

In 4 dimensions, our higher-dimensional 2-charge Kerr--NUT solution is a special case of the NUT generalization of the 4-charge Cveti\v{c}--Youm solution obtained in \cite{crotbh4d}; namely, 2 charge parameters vanish: $\delta_3 = \delta_4 = 0$.  More recently, a 4-dimensional charged Kerr--NUT solution carrying a single charge has been obtained \cite{alcede}.  This 1-charge solution is a special case of the 2-charge solution given above.  To see this, we perform the following coordinate changes on the metric given in \cite{alcede}:
\ben
t = t' + \frac{2 \ell^2 \, \cosh \alpha}{a'} \phi' \, , \quad \phi = \frac{\sqrt{a'^2 + \ell^2}}{a'} \phi' \, , \quad \cos \theta = \frac{y' - \ell}{\sqrt{a'^2 + \ell^2}} \, , \quad a = \sqrt{a'^2 + \ell^2} \, .
\een
Dropping the primes, the metric is $\ud s^2 = (H_1 H_2)^{1/2} \ud \tilde{s}^2$, where $\ud \tilde{s}^2$ is given by (\ref{2chargemetric}) in $D = 4$ with $x_1 = y$, $x_2 = \ui r$, $m_1 = - \ell$, $m_2 = - \ui M$, $\delta_1 = \alpha$, $\delta_2 = 0$.  The other fields also match, after rotating the scalars and normalizing, with the 1-form potentials differing by only an exact form.


\end{document}